\documentclass[journal]{IEEEtran}

\ifCLASSINFOpdf
\else
\fi
\usepackage{amsfonts}   
\usepackage{amsbsy} 
\usepackage{amsmath}
\usepackage{amssymb}
\usepackage{amsthm}
\usepackage{xcolor}
\usepackage{enumitem}

\usepackage{algorithm}
\usepackage{algpseudocode} 

\usepackage{graphicx}
\usepackage{float}
\usepackage{subfloat}
\usepackage[labelformat=simple]{subfig}  

\theoremstyle{plain}
\newtheorem{lemma}{Lemma}
\newtheorem*{remark}{Remark}

\usepackage{cite}



\usepackage{cases}
\usepackage{amsmath,etoolbox}
\patchcmd\subequations
 {\theparentequation\alph{equation}}
 {\subequationsformat}
 {}{}
\newcommand{\subequationsformat}{\theparentequation.\arabic{equation}}

\usepackage{algorithmicx}
\usepackage{algpseudocode}

\usepackage[normalem]{ulem}

\newcommand\mystrut{\rule{0pt}{6.5pt}} 

\begin{document}
%
\title{Robust Filtering and Learning in State-Space Models: Skewness and Heavy Tails Via Asymmetric Laplace Distribution}
%
%
%

\author{Yifan~Yu$^\star$,
        Shengjie~Xiu$^\star$,
        and~Daniel~P.~Palomar,~\IEEEmembership{Fellow,~IEEE}
        \thanks{This work has been submitted to the IEEE for possible publication. Copyright may be transferred without notice, after which this version may no longer be accessible.}
\thanks{This work was supported by the Hong Kong GRF 16206123 research grant. The authors are with the Hong Kong University of Science and Technology (HKUST), Clear Water Bay, Kowloon, Hong Kong (e-mail: yyuco@connect.ust.hk; sxiu@connect.ust.hk; palomar@ust.hk)}\thanks{$^\star$ These authors contributed equally to this work.}}

%
%

\maketitle

\begin{abstract}
State-space models are pivotal for dynamic system analysis but often struggle with outlier data that deviates from Gaussian distributions, frequently exhibiting skewness and heavy tails. This paper introduces a robust extension utilizing the asymmetric Laplace distribution, specifically tailored to capture these complex characteristics. We propose an efficient variational Bayes algorithm and a novel single-loop parameter estimation strategy, significantly enhancing the efficiency of the filtering, smoothing, and parameter estimation processes. 
Our comprehensive experiments demonstrate that our methods provide consistently robust performance across various noise settings without the need for manual hyperparameter adjustments. In stark contrast, existing models generally rely on specific noise conditions and necessitate extensive manual tuning. Moreover, our approach uses far fewer computational resources, thereby validating the model's effectiveness and underscoring its potential for practical applications in fields such as robust control and financial modeling.
\end{abstract}



\section{Introduction} \label{sec: introduction}
State-space models are invaluable tools utilized for describing diverse dynamical systems, with applications spanning robotics, navigation, and financial modeling. These models operate under the assumption that the system's evolution is governed by unobserved hidden states, while only specific measurements can be directly observed. However, a significant challenge in state-space modeling arises from the presence of outliers, which are measurements that deviate markedly from the expected distribution. These outliers can introduce biases and distort the estimation of hidden states, ultimately leading to inaccurate results. For example, unexpected environmental conditions may impact sensor measurements, complicating the accurate determination of the robot's true status.

To effectively address the issue of outliers, there has been a growing interest in extending state-space models to handle non-Gaussian noise distributions that exhibit strong skewness and heavy tails. This extension is crucial for accurately capturing the unique characteristics of outliers in observations and, more importantly, for obtaining a more precise estimation of the hidden states. However, this pursuit comes with three major challenges that must be overcome.

\emph{Challenge 1: Modeling the skewness and heavy tail of outliers.}
Most approaches to noise modeling struggle to efficiently capture the properties of outliers. Early work in outlier modeling focused on non-probabilistic methods, as discussed in \cite{ruckdeschel2014robust, yang2006optimal,chang2013robust}. However, these non-probabilistic methods lack a robust way to quantify the uncertainty associated with outliers and their impact on model estimates. Consequently, they are inflexible for practical use and lack interpretability.
In contrast, probabilistic modeling of outlier distributions provides a mathematical framework to characterize the distribution of outliers \cite{anderson2017black}. Classical heavy-tailed distributions, such as the Student's t-distribution \cite{li2006t, xu2013robust, roth2013student, agamennoni2012approximate} and the Laplace distribution \cite{neri2020laplace, wang2008multivariate,wang2017laplace}, are often considered. However, these methods typically overlook the skewness commonly observed within outliers, thereby resulting in biased estimates. Skewed noise frequently occurs in financial markets where left-tail risk and right-tail risk are prevalent \cite{wang1998actuarial}, and in other applications such as radio signal-based distance estimation \cite{kok2015indoor} and satellite image classification \cite{zadkarami2010application}. 
To address this limitation, researchers have proposed using skewed and heavy-tailed distributions, such as the skew-t distribution, for more robust state-space modeling \cite{Nurminen2015, Nurminen2018}. However, the skew-t distribution presents challenges; its probability density function (PDF) lacks closed-form expressions due to the involvement of special beta functions, and estimating its degrees of freedom is inherently complex \cite{aas2006generalized}, making the model computationally intensive and unstable.
On the other hand, while the asymmetric Laplace (AL) distribution has been primarily utilized in fitting empirical financial data \cite{kozubowski2001asymmetric, trindade2010time}, this paper considers its application to robust state-space modeling \cite{andersson2020optimum}. The AL distribution is chosen for its closed-form density expression, which facilitates easier model evaluation and parameter estimation. By incorporating the AL distribution, we aim to enhance the model’s robustness and computational efficiency, effectively addressing both skewness and heavy-tail characteristics.

\emph{Challenge 2: Efficient inference of the robust state-space models.}
Inference involves estimating hidden states given models with predefined parameters. Particle filtering (PF), also known as the sequential Monte Carlo (SMC) method \cite{doucet2001sequential}, \cite{ bugallo2017adaptive}, is commonly used in robust state-space inference, particularly when dealing with skewed or heavy-tailed distributions. Examples include early PF methods for Student's t and Laplace distributions \cite{li2006t, xu2013robust, wang2008multivariate}, and recent applications to the AL distribution \cite{andersson2020optimum}. However,  these simulation-based methods suffer from a high computational burden as the number of particles required to accurately represent the posterior distribution grows exponentially with the dimension of the states.
An alternative approach is variational Bayes (VB), which provides a deterministic approximation of the posterior distribution, allowing for explicit model solving \cite{marnissi2017variational}. VB has been successfully applied to inference problems involving Student's t distributions \cite{agamennoni2012approximate, agamennoni2011outlier,huang2017novel,fu2022novel}, skew-t distributions \cite{Nurminen2015, Nurminen2018}, and other non-Gaussian distributions \cite{Huangscale,wang2021novel, wang2018robust}. 
Building on this, the VB framework has been extended to Gaussian scale mixtures, which is adaptable to the AL distribution \cite{huang2017robust}. Although this extension broadens VB's applicability, it exhibits notable limitations such as sensitivity to prior distribution choices and the lack of analytical solutions in updates, which may compromise its robustness.

\emph{Challenge 3: Parameter estimation for robust state-space models.}
Another significant issue with the current work is the limited support for parameter estimation in AL-based state-space modeling. Parameter estimation in state-space model is generally more difficult than the inference problem. 
For example, although the mode-based hierarchical approach is efficient for filtering \cite{vila2019decentralized}, it is constrained in parameter estimation because it does not capture the full probabilistic distributions.
Moreover, conventional parameter estimation techniques such as canonical variate analysis \cite{larimore1990canonical} and N4SID \cite{van1993subspace}, which generally assume Gaussian or symmetric noise, are limited in their applicability to scenarios with AL noise \cite{campbell1979canonical}.
The Expectation-Maximization (EM) algorithm is commonly used for parameter estimation in space models, featuring an iterative process that alternates between an E-step and an M-step. Initially, the E-step uses simulation-based methods \cite{durbin1997monte}, which are later replaced by approximate methods to enhance efficiency \cite{neri2020laplace, Huang2016identification}. However, traditional EM approaches are computationally intensive as each iteration requires executing the full smoothing algorithm. To address these challenges, a more efficient EM algorithm specifically tailored for AL-based state-space models is needed.

To address these challenges, this paper makes the following key contributions:
\begin{itemize}
    \item We formulate a robust state-space model using the AL distribution to model measurement noise, providing detailed interpretations that justify its suitability for capturing outlier properties.
    \item We propose a smoothing algorithm and two distinct filtering algorithms, enabling efficient estimation of latent variables within the state-space model.
    \item We develop an efficient novel single-loop parameter estimation algorithm that generalizes and streamlines the variational EM methods.
    \item We present extensive results demonstrating the broader applicability and efficiency of our proposed algorithm compared to existing robust filters. Furthermore, we showcase its effectiveness in practical applications.
\end{itemize}

The rest of the paper is organized as follows. We begin with the robust state-space formulation based on AL distribution in Section \ref{sec:Problem Formulation}. Then, in Section \ref{sec:Filtering and Smoothing}, we propose our filtering and smoothing algorithm based on VB method. The detailed interpretation on the robust effect is provided in Section \ref{sec: interpretation}. Section \ref{sec:Parameter Estimation} provides the fast parameter estimation method for the proposed model. Section \ref{sec: experiments} justifies the proposed algorithm’s broad applicability and performance with comprehensive experiments. Finally, Section \ref{sec: conclusion} concludes this paper.

\section{Problem Formulation} \label{sec:Problem Formulation}
In this section, we begin by formulating the problem of robust state-space modeling with measurement noise characterized by the AL distribution. Subsequently, we explore the capability of the AL distribution to effectively model skewed and heavy-tailed data, which is particularly desirable for handling outliers in measurements.

\subsection{State-Space Model with AL Distributed Noise}
We consider a discrete-time linear state-space model with a hidden state vector $\mathbf{x}_{k} \in \mathcal{R}^{n_x}$ and a measurement vector $\mathbf{y}_{k} \in \mathcal{R}^{n_y}$, where the time step is denoted as $k = 1,\dots, T$. The state-space model is defined as follows:
\begin{equation}
\begin{aligned}  \mathbf{x}_{k+1}&=\mathbf{A}\mathbf{x}_{k}+\mathbf{b}+\mathbf{w}_{k}, & \:& \mathbf{w}_{k}\sim\mathcal{N}\left(\mathbf{0},\mathbf{Q}\right),\\
 \mathbf{y}_{k}&=\mathbf{C}\mathbf{x}_{k}+\mathbf{v}_{k}, & \: & \mathbf{v}_{k}\sim\prod_{i=1}^{n_{y}}\mathcal{AL}\left(\mu_{i},p_{i},\sigma_{i}\right),
\end{aligned}
\label{eq: state space problem}
\tag{$\mathcal{M}$}
\end{equation}
where $\mathbf{A} \in \mathcal{R}^{n_x\times n_x}$ and $\mathbf{C}\in \mathcal{R}^{n_y \times n_x}$ are the state transition and measurement matrices, respectively, $\mathbf{Q}\in \mathcal{R}^{n_x\times n_x}$ is the state noise covariance matrix, and $\mathbf{b} \in \mathcal{R}^{n_x}$ is the bias vector of the state transition. The initial state $\mathbf{x}_1$ follows a Gaussian distribution, i.e., $\mathbf{x}_1 \sim \mathcal{N}(\boldsymbol{\pi}_1, \boldsymbol{\Sigma}_1)$.

The measurement noise vector $\mathbf{v}_{k} \in \mathcal{R}^{n_y}$ is modeled with independent components, each following an asymmetric Laplace distribution. The PDF for a univariate AL-distributed variable $v\sim\mathcal{AL}(\mu, p, \sigma)$ is given by:
\begin{equation}
   f(v) =  \frac{p(1-p)}{\sigma}\exp\left(-\frac{\vert v-\mu \vert  + (2p-1)(v-\mu )}{2\sigma}\right),
\label{eq: ALD PDF}
\end{equation}
where $\mu$ controls the location, $p\in(0,1)$ controls the asymmetry, and $\sigma>0$ controls the scale. In the subsequent sections, we refer to these parameters as  $\boldsymbol{\theta} = \left\{\mathbf{A}, \mathbf{C}, \boldsymbol{\mu}, \mathbf{p}, \boldsymbol{\sigma},\mathbf{b}, \mathbf{Q}, \boldsymbol{\pi}_1, \boldsymbol{\Sigma}_1\right\}$, where $\boldsymbol{\mu}$, $\mathbf{p}$, and $\boldsymbol{\sigma}$ denote the vectors of $\mu_i$, $p_i$, and $\sigma_i$, respectively. For ease of explanation, unless explicitly stated otherwise, we assume the observation dimension $n_y=1$ throughout this paper. Discussions specific to multivariate cases where $n_y>1$ will be clearly highlighted.

This paper addresses two fundamental problems associated with the given state-space model \ref{eq: state space problem}. Firstly, it tackles the challenging task of filtering and smoothing, which involves accurately estimating the hidden state vector $\mathbf{x}_{k}$ given known parameters and observations. Secondly, it focuses on the parameter estimation problem, aiming to robustly estimate the unknown model parameters based on the available data.

\subsection{Motivation of Using the AL Distribution\label{subsec:motivation AL}}

\begin{figure}[]
\begin{centering}
\subfloat[]{\includegraphics[width=0.5\linewidth]{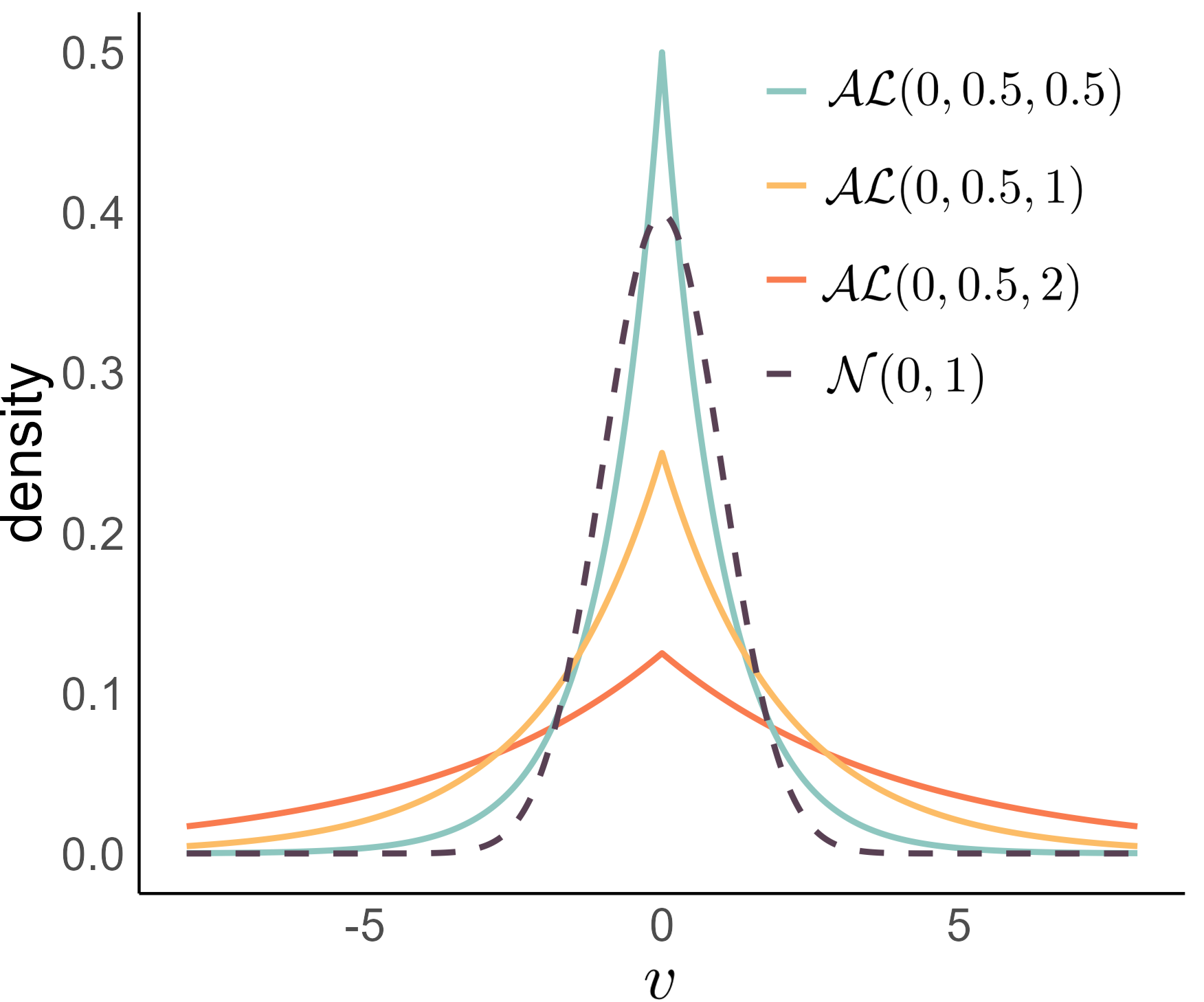}\label{fig: pdf-varying sigma}}
\subfloat[]{\includegraphics[width=0.5\linewidth]{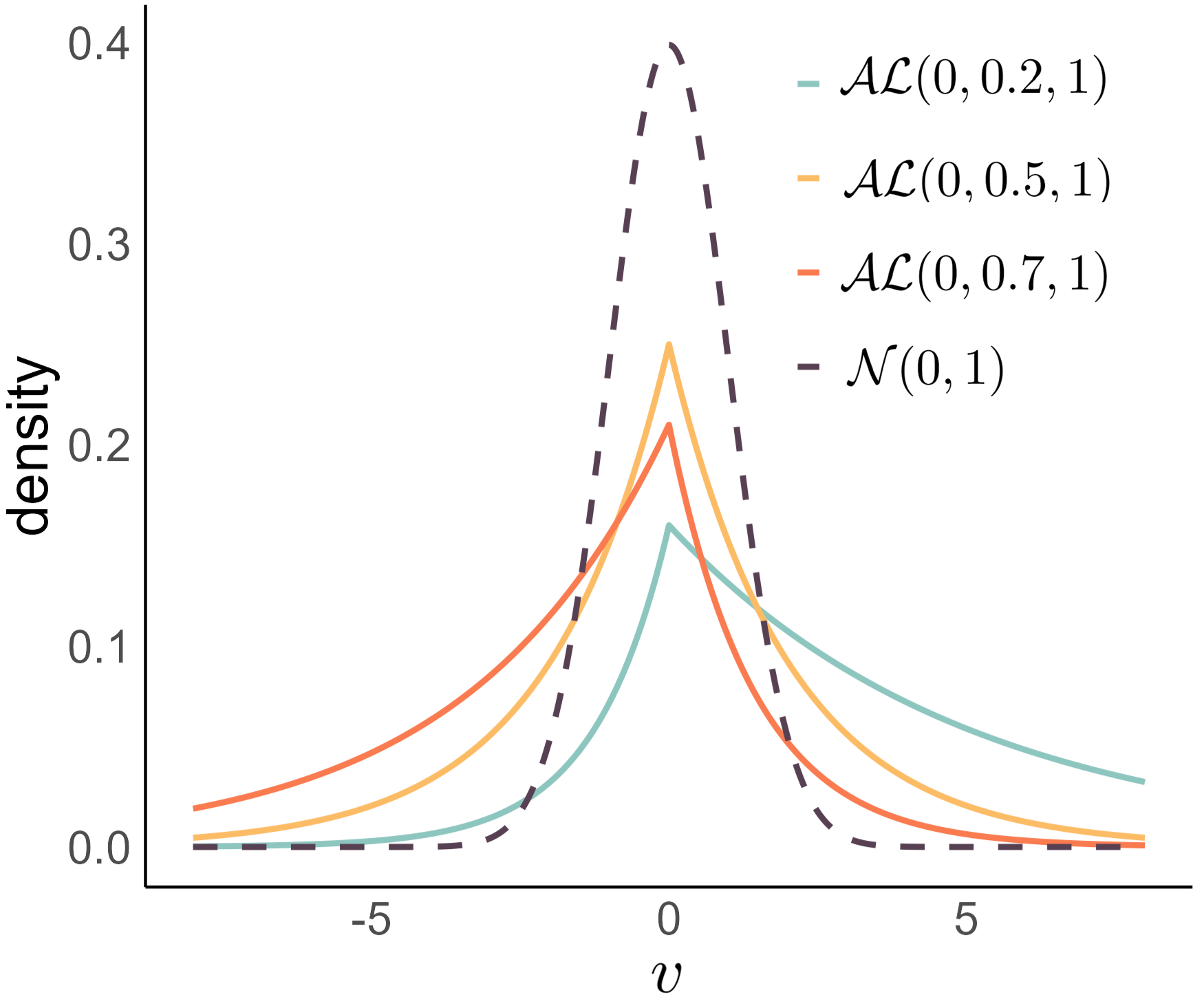}\label{fig: pdf-varying p}}
\par\end{centering}
\caption{The PDF $\mathcal{AL}(0,p,\sigma)$ for different $p$ and $\sigma$ in comparison with $\mathcal{N}(0,1)$.}
\label{fig:pdf}
\end{figure}

In this subsection, we explore how the AL distribution effectively models the skewed and heavy-tailed properties of outliers, addressing the first challenge discussed in Section \ref{sec: introduction}.
As shown in Fig. \ref{fig: pdf-varying sigma}, the AL distribution (with $p=0.5$) exhibits heavier tails than the Gaussian distribution, indicating a higher probability of extreme events. Furthermore, increasing the parameter $\sigma$ results in even fatter tails.
Fig. \ref{fig: pdf-varying p} illustrates the skewness of the AL distribution as $p$ deviates from 0.5. When $p$ approaches 0, the distribution becomes positively skewed, indicating a higher likelihood of positive outliers. Conversely, as $p$ approaches 1, the skewness reverses, favoring negative outliers.

\begin{figure}[]
    \centering
    \includegraphics[width=0.9\linewidth]{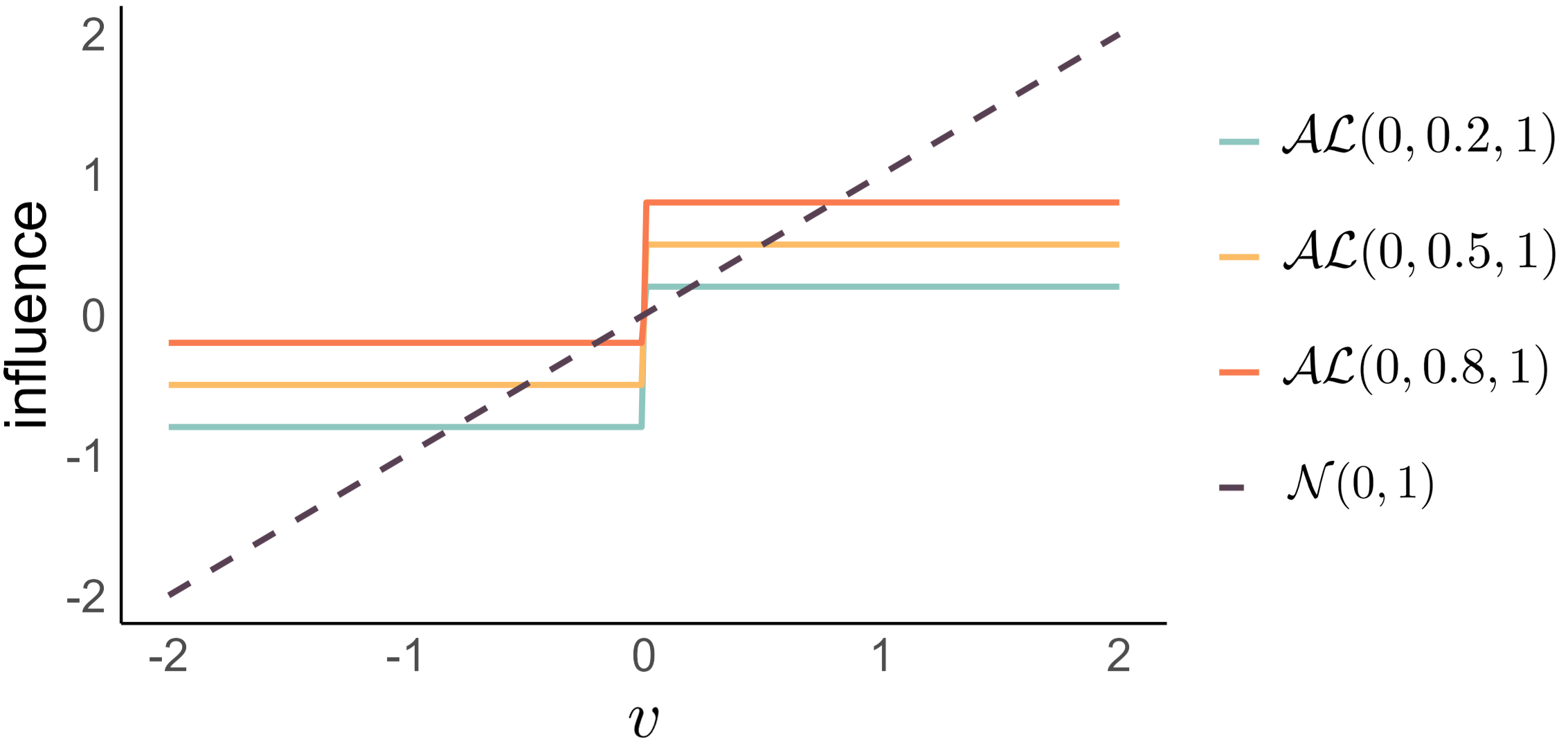}
    \caption{Influence function of $\mathcal{AL}(0,p,1)$ and $\mathcal{N}(0,1)$.}
    \label{fig:influence fun}
\end{figure}

To assess the AL distribution's robustness against outliers compared to the Gaussian model, we examine their influence functions \cite{hampel1986influence}, defined as ${d \log f(v)}/{d v}$, as depicted in Fig. \ref{fig:influence fun}. This function illustrates how minor changes in data affect measurement likelihoods, which indirectly affect state and parameter estimation in state-space models. Under Gaussian assumptions, the sensitivity to outliers increases as the noise magnitude increases, potentially destabilizing the model. Conversely, the AL distribution’s bounded influence function enhances robustness by limiting the impact of extreme outliers. At $p = 0.2$, suggesting a heavier right tail, the AL model responds more strongly to negative than to positive outliers, showing greater tolerance for positive deviations and effectively managing right-skewed data. Similarly, with $p > 0$, it adapts to handle left-skewed data efficiently.

In Section \ref{sec:Filtering and Smoothing} and \ref{sec: interpretation}, we will provide detailed insights into how incorporating the AL distribution as the measurement noise in the state-space model \ref{eq: state space problem} enables robustness and accurate estimation of the system state.

\section{Smoothing and Filtering Algorithms}\label{sec:Filtering and Smoothing}
In this section, we present the derivation of the smoothing and filtering algorithms for the proposed state-space model with AL distributed measurement noise. For clarity in subsequent discussions, we introduce key notation early. The variables $\hat{\mathbf{x}}_{k\vert k-1}$, $\hat{\mathbf{x}}_{k\vert k}$, and $\hat{\mathbf{x}}_{k\vert T}$ denote the mean estimates for the predicted, filtered, and smoothed states, respectively. Similarly, the matrices $\boldsymbol{\Sigma}_{k\vert k-1}$, $\boldsymbol{\Sigma}_{k\vert k}$, and $\boldsymbol{\Sigma}_{k\vert T}$ represent the covariance matrices of these estimates at the respective stages, quantifying the uncertainty associated with each estimate under the Gaussian assumptions.

Both algorithms are based on the hierarchical representation of the AL distribution given in the following lemma.
\begin{lemma}
    AL distribution can be represented through a hierarchical model involving Gaussian and Inverse-Gamma distributions by introducing an auxiliary variable $\lambda$ \cite{wand2011mean}. Given $v \sim \mathcal{AL}(\mu, p, \sigma)$, we have 
\begin{equation}
     v|\lambda \sim \mathcal{N}\left(\mu + \frac{(\frac{1}{2}-p)\sigma}{\lambda p(1-p)},\frac{\sigma^2}{\lambda p(1-p)}\right),
\end{equation}
\begin{equation}
    \lambda \sim \text{Inverse-Gamma}\left(1,\frac{1}{2}\right).
\end{equation}
\label{lemma: ALD hierachy}
\end{lemma}
Based on Lemma \ref{lemma: ALD hierachy}, the measurement equation of  the proposed state-space model \ref{eq: state space problem} can be represented as:
\begin{equation}
     y_{k}\vert\mathbf{x}_{k}, \lambda_{k} \sim \mathcal{N}\left(\mathbf{C}\mathbf{x}_{k} +\mu+ \frac{(\frac{1}{2}-p)\sigma}{\lambda_{k} p(1-p)}, \frac{\sigma^2}{\lambda_{k} p(1-p)}\right),
\label{eq: prob y given x}
\end{equation}
\begin{equation}
    \lambda_{k} \sim \text{Inverse-Gamma}\left(1,\frac{1}{2}\right).
\label{eq: prob lambda}
\end{equation}
Here, $\lambda_k \in \mathcal{R}$ is an auxiliary variable updated at each time step, representing the adaptive time-variant scale of the variance.

\subsection{Smoothing Algorithm Based on MFVB}\label{subsec: smoothing algorithm}
Smoothing is a crucial technique in state estimation that involves estimating past hidden states given the complete sequence of observations. The objective of the smoothing algorithm is to estimate the joint posterior distribution $P(\mathbf{x}_{1:T}, \lambda_{1:T} \vert y_{1:T})$, which represents the distribution of all hidden states $\mathbf{x}_{1:T}$ and auxiliary variables $\lambda_{1:T}$ conditioned on the complete observation sequence $y_{1:T}$. In the MFVB framework, this joint posterior is approximated by a factorized distribution:
\begin{equation}
    P\left(\mathbf{x}_{1:T}, \lambda_{1:T} \vert y_{1:T}\right) \approx q_x\left(\mathbf{x}_{1:T}\right)q_{\lambda}\left(\lambda_{1:T}\right).
\label{eq: MFVB smoother}
\end{equation}

To minimize the Kullback–Leibler (KL) divergence between these variational distributions and the true posterior for the best approximation, we iteratively update $q_{x}(\cdot)$ and $q_{\lambda}(\cdot)$ using the formulas provided in \cite{bishop2006pattern}:
\begin{subequations}
\small
\begin{align}
    \log q_{x}\left(\mathbf{x}_{1:T}\right) &= \mathbb{E}_{q_{\lambda}}\left[\log P\left(\mathbf{x}_{1:T}, \lambda_{1:T}, y_{1:T}\right)\right] + \text{const}, \label{eq: smoother update of q_x}\\
    \log q_{\lambda}\left(\lambda_{1:T}\right) &= \mathbb{E}_{q_x}\left[\log P\left(\mathbf{x}_{1:T}, \lambda_{1:T}, y_{1:T}\right)\right] + \text{const}. \label{eq: smoother update of q_lambda}
\end{align}
\end{subequations}

The iterative updates at the $j$-th iteration proceed as follows, with detailed derivations provided in Appendix \ref{appendix:update_x_and_lambda}.

\subsubsection*{Update of $q_{x}^{j+1}(\mathbf{x}_{1:T})$}

Given $q^{j}_\lambda(\lambda_{1:T})$, the variational posterior $q^{j+1}_{x}(\mathbf{x}_{1:T})$ takes the form:
\begin{equation}
\label{eq:q_x_1_T closed-form}
q^{j+1}_{x}\left(\mathbf{x}_{1:T}\right)\propto P(\mathbf{x}_1)\prod_{k = 2}^T P(\mathbf{x}_k\vert \mathbf{x}_{k-1})\prod_{k= 1}^T P\left(y_k\vert \mathbf{x}_k; \mathbb{E}_{q_{\lambda}^{j}}[\lambda_{k}]\right),
\end{equation}
where 
\begin{equation}
\begin{aligned}
       &P\left(y_k\vert \mathbf{x}_k; \mathbb{E}_{q_{\lambda}^{j}}[\lambda_{k}]\right)\\ = &\mathcal{N}\left(\mathbf{Cx}_k + \frac{\left(\frac{1}{2}-p\right)\sigma}{\mathbb{E}_{q_{\lambda}^{j}}[\lambda_{k}]p\left(1-p\right)},\frac{\sigma^{2}}{\mathbb{E}_{q_{\lambda}^{j}}[\lambda_{k}]p\left(1-p\right)}\right).
\end{aligned}
\end{equation}
This reduces to the smoothing problem for a Gaussian system with time-varying measurement noise parameters.

\subsubsection*{Update of $q^{j+1}_{\lambda}(\lambda_{1:T})$}
Given $q^{j+1}_x(\mathbf{x}_{1:T})$, the variational posterior $q^{j+1}_{\lambda}(\lambda_{1:T})$ takes the form:
\begin{equation}
\label{eq:smoother q_lambda closed-form}
\small
q^{j+1}_{\lambda}\left(\lambda_{1:T}\right)=\prod_{k = 1}^T\mathcal{IG}\left(\frac{\sigma}{2p(1-p)\sqrt{u_k^{j+1}}},\frac{1}{4p(1-p)}\right),
\end{equation}
where $u_k^{j+1} = \mathbb{E}_{q_x^{j+1}}[(y_{k}-\mathbf{C}\mathbf{x}_{k}-\mu)^{2}]$.

The proposed, as detailed in Algorithm \ref{al:smoother}, exhibits structural similarities with the traditional Kalman smoother, most notably in its adoption of a forward-backward procedure. However, in contrast to the Kalman smoother, which typically requires only a single forward and backward pass, our proposed smoother necessitates multiple iterations. This iterative approach is crucial for refining the estimation of the auxiliary variables \(\lambda_{1:T}\), which address the non-Gaussian characteristics of the measurement noise. 

\begin{algorithm}[t]
\caption{AL-Smoother for the state-space model \ref{eq: state space problem}\protect\label{al:smoother}}
    \begin{flushleft}
    \textbf{Input:} $\boldsymbol{\theta}$, $y_{1:T}$, initial guess of $\mathbb{E}[\lambda_{k}]>0$ for $k=2,\dots,T$.
    \end{flushleft}
   \begin{algorithmic}[1]
    \State Let $\hat{\mathbf{x}}_{1\vert 1} = \boldsymbol{\pi}_1$, $\boldsymbol{\Sigma}_{1\vert1} = \boldsymbol{\Sigma}_1$
    \Repeat
    \For{$k =2:T$} 
    \State $\hat{\mathbf{x}}_{k|k-1} = \mathbf{A}\hat{\mathbf{x}}_{k-1|k-1}+ \mathbf{b}$
    \State $\boldsymbol{\Sigma}_{k|k-1} = \mathbf{A}\boldsymbol{\Sigma}_{k-1|k-1}\mathbf{A}\mystrut^\intercal + \mathbf{Q}$
    \State $r_k = \displaystyle{\frac{\sigma^2}{\mathbb{E}_{q_\lambda}[\lambda_{k}] p(1-p)}}$
    \State $m_k =\mu + \displaystyle{ \frac{(\frac{1}{2}-p)\sigma}{\mathbb{E}_{q_\lambda}[\lambda_{k}] p(1-p)}} $
    \State $\mathbf{K}_{k} = \boldsymbol{\Sigma}_{k|k-1} \mathbf{C}\mystrut^\intercal\left(\mathbf{C}\boldsymbol{\Sigma}_{k|k-1}\mathbf{C}\mystrut^\intercal + r_k\right)^{-1}$
    \State $\hat{\mathbf{x}}_{k|k} = \hat{\mathbf{x}}_{k|k-1} + \mathbf{K}_{k}\left(y_{k}- \mathbf{C}\hat{\mathbf{x}}_{k|k-1} -m_k\right)$
    \State $\boldsymbol{\Sigma}_{k|k} = \boldsymbol{\Sigma}_{k|k-1} - \mathbf{K}_{k}\mathbf{C}\boldsymbol{\Sigma}_{k|k-1}$
    \EndFor
    \For{$k =T-1:1$} 
    \State $\mathbf{L}_k = \boldsymbol{\Sigma}_{k|k}\mathbf{A}\mystrut^\intercal\boldsymbol{\Sigma}_{k+1|k}^{-1}$
    \State $\hat{\mathbf{x}}_{k|T} = \hat{\mathbf{x}}_{k|k} + \mathbf{L}_k\left(\hat{\mathbf{x}}_{k+1|T} - \hat{\mathbf{x}}_{k+1|k}\right)$
    \State $\boldsymbol{\Sigma}_{k|T} = \boldsymbol{\Sigma}_{k|k} + \mathbf{L}_k\left(\boldsymbol{\Sigma}_{k+1|T} - \boldsymbol{\Sigma}_{k+1|k}\right)\mathbf{L}\mystrut^\intercal_k$
    \EndFor
    \For{$k =1:T$} 
    
    \State $u_k = \left(y_{k}-\mathbf{C}\hat{\mathbf{x}}_{k|T} -\mu\right)^2 + \mathbf{C}\boldsymbol{\Sigma}_{k\vert T}\mathbf{C}\mystrut^\intercal$
    \State $\mathbb{E}_{q_\lambda}[\lambda_{k}] = \displaystyle{\frac{\sigma}{2p(1-p)\sqrt{u_k}}}$
    \EndFor
    \Until{$\|\Delta\hat{\mathbf{x}}_{1:T|T}+\Delta\text{vec}(\boldsymbol{\Sigma}_{1:T|T})\| <$ threshold}
	\end{algorithmic}
    \begin{flushleft}
    \textbf{Output:} $\hat{\mathbf{x}}_{k|T},\boldsymbol{\Sigma}_{k|T}$ for $k = 1, \dots, T$.
    \end{flushleft}
\end{algorithm}

\subsection{Filtering Algorithm Based on MFVB}
Filtering involves estimating the hidden states $\mathbf{x}_k$ based on observations $y_{1:k}$. In Gaussian systems, this involves Bayesian updating of the posterior state estimate by combining the state prediction with the current measurement:
\begin{equation}
    P\left(\mathbf{x}_{k}\vert y_{1:k}\right)\propto P\left(\mathbf{x}_{k}\vert y_{1:k-1}\right)P\left(y_{k}\vert\mathbf{x}_{k}\right),
\label{eq:Bayes rule}
\end{equation}
where $P\left(\mathbf{x}_{k}\vert y_{1:k-1}\right)$ is Gaussian.

However, for non-Gaussian systems, the predictive distribution $P\left(\mathbf{x}_{k}\vert y_{1:k-1}\right)$, which is a posterior of $\mathbf{x}_k$ based on all observations up to time $k-1$, becomes analytically intractable. To address this fundamental challenge, we develop two distinct filtering algorithms with different computational-accuracy trade-offs.

\subsubsection{Exact AL-Filter}

One approach to handle the intractable predictive distribution is through batch estimation, which computes the joint posterior $P(\mathbf{x}_{1:k} \vert y_{1:k})$ for the entire state sequence up to time $k$. This approach leverages the AL-Smoother framework applied to progressively expanding observation windows, as formalized in Algorithm \ref{al:VB-filter}.
\begin{algorithm}[t]
\caption{Exact AL-Filter for the state-space model \ref{eq: state space problem}}\label{al:VB-filter}
    \begin{flushleft} 
    \textbf{Input:} $\boldsymbol{\theta}$, $y_{1:T}$, initial guess of $\mathbb{E}[\lambda_{k}]>0$ for $k=2,\dots,T$.
    \end{flushleft}
   \begin{algorithmic}[1]
   \For{$k =1:T$} 
    \State Apply AL-Smoother (Algorithm \ref{al:smoother}) to observations $y_{1:k}$ to obtain $\hat{\mathbf{x}}_{k|k},\boldsymbol{\Sigma}_{k|k}$.
    \EndFor
	\end{algorithmic}
    \begin{flushleft}
    \textbf{Output:} $\hat{\mathbf{x}}_{k|k},\boldsymbol{\Sigma}_{k|k}$ for $k = 1, \dots, T$.
    \end{flushleft}
\end{algorithm}

While this method provides rigorous treatment within the VB framework, the computational cost increases significantly with sequence length, as each update requires reprocessing the entire observation history. This motivates the development of a more efficient alternative.

\subsubsection{Fast AL-Filter}
For real-time applications where computational efficiency is paramount, we develop a sequential filtering algorithm that maintains the VB framework while introducing a tractable approximation for the predictive distribution. The key insight is to replace the intractable predictive distribution with a Gaussian approximation derived from the previous time step's variational posterior.

Following the VB framework, we factorize the joint posterior distribution as $P(\mathbf{x}_k, \lambda_k \vert y_{1:k}) \approx q_x(\mathbf{x}_k)q_{\lambda}(\lambda_k)$. This factorization leads to the following iterative update steps:
\begin{subequations}
\begin{align}
    \log q_{x}\left(\mathbf{x}_{k}\right) &= \mathbb{E}_{q_\lambda}\left[\log P\left(\mathbf{x}_{k}, \lambda_{k}, y_{1:k}\right)\right] + \text{const}, \label{eq: update of f q_x}\\
    \log q_{\lambda}\left(\lambda_{k}\right) &= \mathbb{E}_{q_\mathbf{x}}\left[\log P\left(\mathbf{x}_{k}, \lambda_{k}, y_{1:k}\right)\right] + \text{const},\label{eq: update of f q_lambda}
\end{align}
\end{subequations}
where 
\begin{equation}
\begin{aligned}
        &P\left(\mathbf{x}_{k}, \lambda_{k}, y_{1:k}\right) \\
        = &P\left(\mathbf{x}_{k}|y_{1:k-1}\right)P\left(\lambda_{k}\right)P\left(y_{k}|\mathbf{x}_{k},\lambda_{k}\right)P\left(y_{1:k-1}\right).
\end{aligned}
\end{equation}
The computational bottleneck arises from the intractable term $P\left(\mathbf{x}_{k}|y_{1:k-1}\right)$. Rather than resorting to sampling-based methods, we employ analytical approximations for computational efficiency. To achieve computational tractability, we employ the Gaussian approximation:
\begin{equation}
\mathbf{x}_k \vert y_{1:k-1} \sim \mathcal{N}(\mathbf{A}\hat{\mathbf{x}}_{k-1\vert k-1} +\mathbf{b}, \mathbf{A}\boldsymbol{\Sigma}_{k-1 \vert k-1}\mathbf{A}^\top + \mathbf{Q})
\end{equation}
derived from the previous time step's variational posterior. This approximation enables closed-form updates while maintaining the essential structure of the VB framework. The iterative updates at the $j$-th iteration proceed as follows, with detailed derivations provided in Appendix \ref{appendix:filter update_x_and_lambda}.

\subsubsection*{Update of $q^{j+1}_{x}(\mathbf{x}_{k})$}
Given the conjugate Gaussian prior, the variational posterior $q^{j+1}_{x}(\mathbf{x}_{k})$ takes a Gaussian form:
\begin{equation}
\label{eq:filter q_x closed-form}
q^{j+1}_{x}\left(\mathbf{x}_{k}\right)=\mathcal{N}\left(\hat{\mathbf{x}}_{k\vert k},\boldsymbol{\Sigma}_{k\vert k}\right).
\end{equation}

\subsubsection*{Update of $q^{j+1}_{\lambda}(\lambda_{k})$}
Using the conjugate prior $P(\lambda_k) = \text{Inverse-Gamma}(1, \frac{1}{2})$, which is a special case of the generalized inverse Gaussian distribution, the posterior $q^{j+1}_{\lambda}(\lambda_{k})$ follows the inverse Gaussian distribution:
\begin{equation}
\label{eq:filter q_lambda closed-form}
\small
q^{j+1}_{\lambda}\left(\lambda_{k}\right)=\mathcal{IG}\left(\frac{\sigma}{2p(1-p)\sqrt{u_k^{j+1}}},\frac{1}{4p(1-p)}\right),
\end{equation}
where $u_k^{j+1} = \mathbb{E}_{q_x^{j+1}}[(y_{k}-\mathbf{C}\mathbf{x}_{k}-\mu)^{2}]$.

The resulting filter algorithm (Algorithm \ref{al:fast filter}) maintains the Kalman filter framework, with an unchanged time predict step but a looped measurement update due to the cyclic approximation of the posterior.
Extension to multivariate cases assumes independent measurement noise elements, yielding:
\begin{equation}
    P\left(\mathbf{x}_{k}, \boldsymbol{\lambda}_{k} \vert \mathbf{y}_{1:k}\right) \approx q_{x}\left(\mathbf{x}_{k}\right) \prod_{i=1}^{n_{y}}q_{\lambda}\left(\lambda_{k,i}\right).
\end{equation}
Each $\lambda_{k,i}$ update follows Equation \eqref{eq:filter q_lambda closed-form}. Incorporating these updates into Algorithm \ref{al:fast filter} involves replacing $r_k$ and $m_k$ with $\mathbf{R}_k = \text{diag}(r_{k,1},\dots,r_{k,n_y})$ and $\mathbf{m}_k = [m_{k,1},\dots, m_{k,n_y}]^\intercal$. 

\begin{algorithm}[t]
\caption{Fast AL-Filter for the state-space model \ref{eq: state space problem}\protect\label{al:fast filter}}
    \begin{flushleft} 
    \textbf{Input:} $\boldsymbol{\theta}$, $y_{1:T}$, initial guess of $\mathbb{E}[\lambda_{k}]>0$ for $k=2,\dots,T$.
    \end{flushleft}
    \begin{algorithmic}[1]
 \State Let $\hat{\mathbf{x}}_{1|1} = \boldsymbol{\pi}_1$, ${\Sigma}_{1|1} = \boldsymbol{\Sigma}_1$
 \For{$ k = 2, \dots, T$}
 \State $\hat{\mathbf{x}}_{k|k-1} = \mathbf{A}\hat{\mathbf{x}}_{k-1|k-1} + \mathbf{b}$
 \State $\boldsymbol{\Sigma}_{k|k-1} = \mathbf{A}\boldsymbol{\Sigma}_{k-1|k-1}\mathbf{A}^\intercal + \mathbf{Q}$
    \Repeat
    \State $r_k = \displaystyle{\frac{\sigma^2}{\mathbb{E}_{q_\lambda}[\lambda_{k}] p(1-p)}}$
    \State $m_k = \mu + \displaystyle{\frac{(\frac{1}{2}-p)\sigma}{\mathbb{E}_{q_\lambda}[\lambda_{k}] p(1-p)}}$
    \State $\mathbf{K}_k = \boldsymbol{\Sigma}_{k|k-1} \mathbf{C}\mystrut^\intercal\left(\mathbf{C}\boldsymbol{\Sigma}_{k|k-1}\mathbf{C}\mystrut^\intercal + r_k\right)^{-1}$
    \State $\hat{\mathbf{x}}_{k|k} = \hat{\mathbf{x}}_{k|k-1} + \mathbf{K}_k\left(y_{t}- \mathbf{C}\hat{\mathbf{x}}_{k|k-1} - m_k\right)$
    \State $\boldsymbol{\Sigma}_{k|k} = \boldsymbol{\Sigma}_{k|k-1} - \mathbf{K}_k\mathbf{C}\boldsymbol{\Sigma}_{k|k-1}$
    \State $u_k = \left(y_{k}-\mathbf{C}\hat{\mathbf{x}}_{k|k} -\mu\right)^2 + \mathbf{C}\boldsymbol{\Sigma}_{k\vert k}\mathbf{C}\mystrut^\intercal$
    \State $\mathbb{E}_{q_\lambda}[\lambda_{k}] = \displaystyle{\frac{\sigma}{2p(1-p)\sqrt{u_k}}}$
    \Until{$\|\Delta\hat{\mathbf{x}}_{k|k}+\Delta\text{vec}(\boldsymbol{\Sigma}_{k|k})\| <$ threshold}
    \EndFor
	\end{algorithmic}
    \begin{flushleft}
    \textbf{Output:} $\hat{\mathbf{x}}_{k|k-1},\boldsymbol{\Sigma}_{k|k-1}, \hat{\mathbf{x}}_{k|k}, \boldsymbol{\Sigma}_{k|k}$ for $k = 2,\dots,T$.
    \end{flushleft}
\end{algorithm}

\section{Interpretation of Robustness in Proposed State-Space Model}\label{sec: interpretation}

This section delves into the robustness by examining it from two distinct perspectives. The latter is specific to the Fast AL-Filter, which incorporates an additional Gaussian assumption on the predictive distribution. To facilitate this exploration, we consider a simplified model \ref{eq: simple state space illustration}, where $n_x = 1$ and $n_y = 1$, with  $A=C=1$ and $b=0$:
\begin{equation}
\begin{aligned}  {x}_{k+1}&={x}_{k}+{w}_{k}, & \quad & {w}_{k}\sim\mathcal{N}\left(0,q\right),\\
 {y}_{k}&={x}_{k}+{v}_{k}, & \quad & {v}_{k}\sim\mathcal{AL}\left(\mu,p,\sigma\right).
\end{aligned}
\tag{$\mathcal{M}_s$}
\label{eq: simple state space illustration}
\end{equation}
At time $k$, given the prior $P(x_k \vert y_{1:k-1}) = \mathcal{N}(\hat{x}_{k\vert k-1}, \Sigma_{k\vert k-1})$ and varying noise values $v_k$ in the observation, we analyze the robustness of filters in estimating the state $\hat{x}_{k|k}$.

\subsection{Viewpoint 1: Adaptive Filtering Approach\label{subsec: adaptive filter}}
Adaptive filtering dynamically adjusts filter parameters in response to changing noise characteristics and system dynamics. This technique is crucial for managing outliers in state estimation, particularly when dealing with non-stationary or complex noise characteristics. One common non-probabilistic technique \cite{mohamed1999adaptive} updates the noise covariance matrix using past innovations $d_k = y_k - \mathbf{A}\hat{\mathbf{x}}_{k\vert k-1}$ as:
\begin{equation}
\label{eq: adaptive R in mohamed}
    r_k = \frac{1}{N_{\text{win}}}\sum_{i = 1}^{N_{\text{win}}} d_{k-i}^2 - \mathbf{C}\boldsymbol{\Sigma}_{k\vert k-1}^{-1}\mathbf{C}\mystrut^\intercal,
\end{equation}
where $N_{\text{win}}$ is the length of the look-back window. However, selecting an appropriate $N_{\text{win}}$ involves a trade-off: a small $N_{\text{win}}$ can make the filter overly sensitive to recent changes, leading to rapid but potentially noisy adjustments; a large $N_{\text{win}}$ smooths out fluctuations, providing stability but possibly delaying the filter's response to sudden changes. This process often requires manual tuning and lacks clear technical guidelines.

\begin{figure}[]
    \centering
    \includegraphics[width=0.9\linewidth]{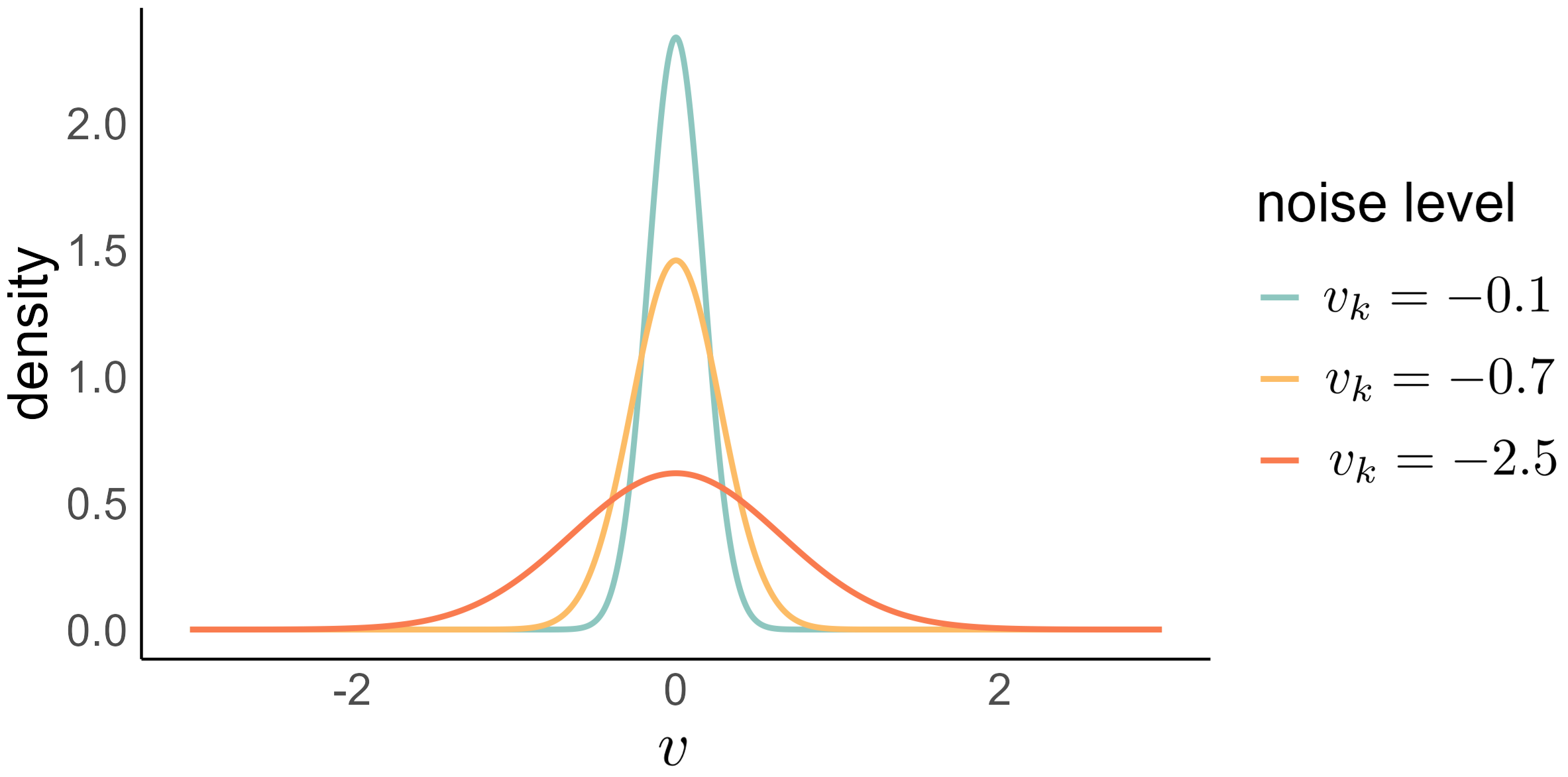}
    \caption{PDF of the noise variable $\hat{v}_k$ estimated by Algorithm \ref{al:fast filter} under different noise levels $v_k$.}
    \label{fig:y-distribution}
\end{figure}

In contrast, our proposed model is a probabilistic technique that leverages the AL distribution to dynamically adapt to the skewness and heavy tail within the measurement noise, eliminating the need for manual parameter tuning while exhibiting stronger robustness. By introducing the auxiliary variable $\lambda_k$, the AL-distributed noise is equivalent to an adaptive Gaussian noise with variance:
\begin{equation}
\label{eq:adaptive noise}
r_{k}=\frac{\sigma^{2}}{\mathbb{E}_{q_\lambda}\left[\lambda_{k}\right]p(1-p)}=2\sigma\sqrt{\mathbb{E}_{q_x}\left[\left(y_{k}-\mathbf{C}\hat{\mathbf{x}}_{k\vert k}-\mu\right)^{2}\right]}.
\end{equation}
This formulation enables $r_k$ to increase when there are significant deviations between the observed and predicted measurements, enhancing the model's adaptability to outlier effects. Fig. \ref{fig:y-distribution} shows how the PDF of the estimated measurement noise variable $\hat{v}_k$, denoted as $\hat{v}_k \sim P(v\vert x_k, \lambda_k, y_k)$, adapts with increasing noise magnitude.

Furthermore, our model effectively handles skewed noise distributions by dynamically updating the mean of the measurement noise:
\begin{equation}
\label{eq: adaptive mean}
\begin{aligned}m_{k} & =\mu+\frac{(\frac{1}{2}-p)\sigma}{\mathbb{E}_{q_\lambda}\left[\lambda_{k}\right]p(1-p)}\\
 & =\mu+(1-2p)\sqrt{\mathbb{E}_{q_x}\left[\left(y_{k}-\mathbf{C}\hat{\mathbf{x}}_{k\vert k}-\mu\right)^{2}\right]}.
\end{aligned}
\end{equation}
The term $(1-2p)$ adjusts for skewness. When $p<0.5$, indicating right-skewness, $m_k$ increases above $\mu$ to account for the predominance of the rightward tail, and vice versa.  This targeted adjustment helps mitigate the bias that skewness in noise distributions might introduce, enhancing the model’s alignment with real-world data characteristics and improving estimation accuracy.

\subsection{Viewpoint 2: Bayesian Robustness Interpretation for the Fast AL-Filter\label{subsec:bayesian robustness}}
When estimating the hidden state $\hat{x}_{k|k}$ via filtering, two sources of information are utilized: predicted state $\hat{x}_{k|k-1}$ and current observations $y_k$, as shown in Equation \eqref{eq:Bayes rule}. In the presence of outliers in the measurement, observations that significantly deviate from their predicted values should be treated with caution and assigned less weight in the posterior distribution \cite{meinhold1989robustification}.

We analyze the robustness of our proposed algorithm from two perspectives: exact Bayesian inference and VB inference. The exact Bayesian perspective offers a theoretical foundation, while the VB perspective demonstrates practical performance.

\subsubsection{Exact Bayesian inference} From the exact Bayesian perspective, the maximum a posteriori probability (MAP) estimate of $\hat{x}_{k|k}$ is obtained by:
\begin{equation}
\label{eq:bayesian robust}
    \hat{x}_{k\vert k} = \hat{x}_{k\vert k-1} + \Sigma_{k\vert k-1}\left.\frac{d\log  g_k(x)}{d x}\right\vert_{x = \hat{x}_{k\vert k-1}} .
\end{equation}
The formulation of $g_k(\cdot)$ and a detailed derivation is provided in Appendix \ref{appendix:simple model}.

\begin{figure}[]
    \centering
    \includegraphics[width=0.9\linewidth]{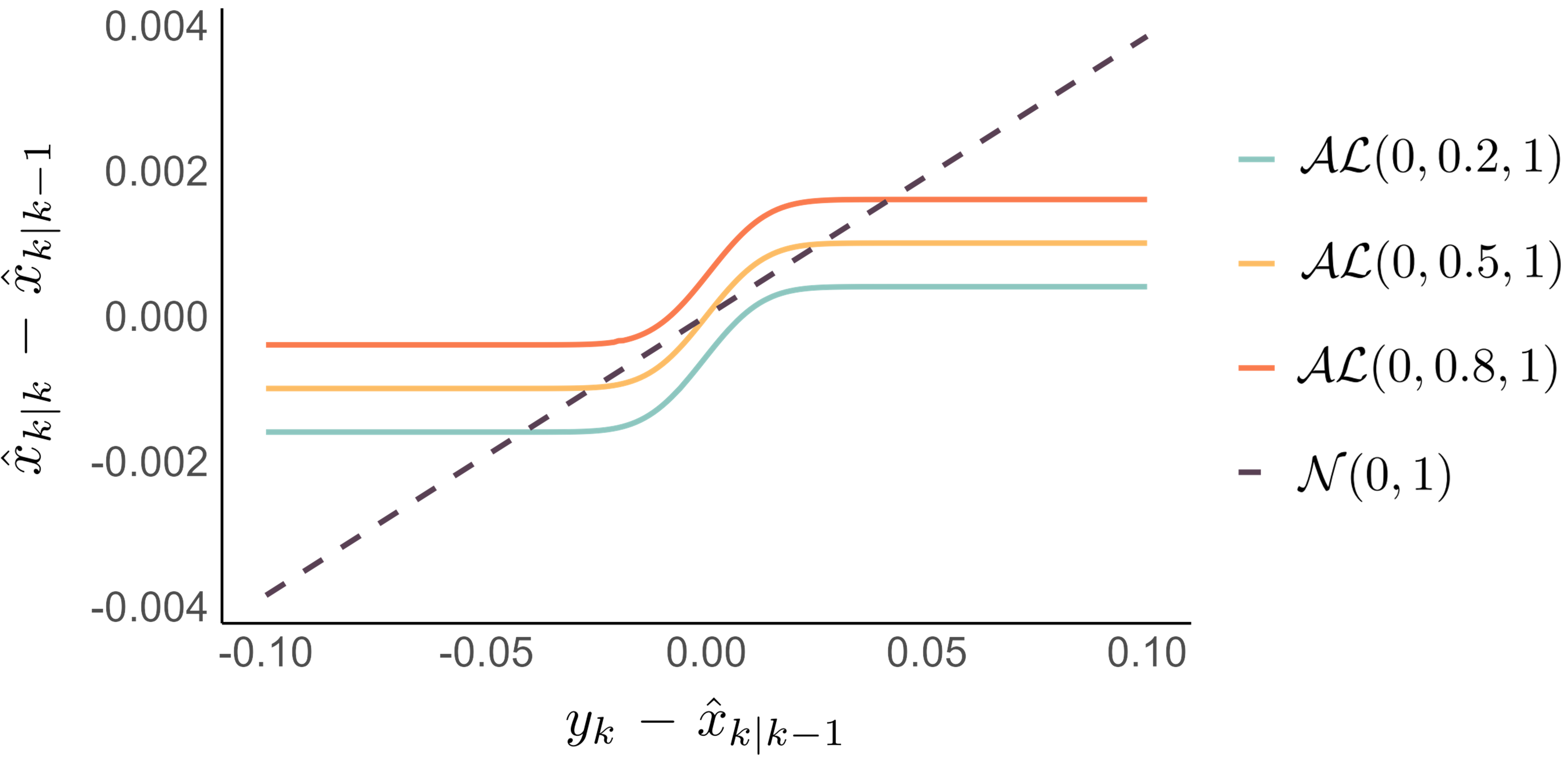}
    \caption{Changes in $\hat{x}_{k \vert k} - \hat{x}_{k \vert k-1}$ against $y_k - \hat{x}_{k \vert k-1}$ for different noise assumptions.}
    \label{fig:post mean}
\end{figure}
Based on Equation \eqref{eq:bayesian robust}, Fig. \ref{fig:post mean} illustrates how the update of the state estimate $\hat{x}_{k \vert k} - \hat{x}_{k \vert k-1}$ responds to the innovation $y_k - \hat{x}_{k \vert k-1}$ under exact Bayesian inference for different noise assumptions. Under the Gaussian noise assumption, the magnitude of $\hat{x}_{k \vert k}$ increases significantly as the innovation increases. In contrast, under the AL distributed noise assumption, when the innovation is high, indicating a potential outlier, its influence on $\hat{x}_{k \vert k}$ is limited. This behavior highlights the robustness of the AL distribution in managing outliers, ensuring that extreme values do not overly influence the state estimation. This aligns with the motivations discussed in Section \ref{subsec:motivation AL}.

\begin{figure}
    \centering    
    \subfloat[Kalman filter]{\includegraphics[width=1\linewidth]{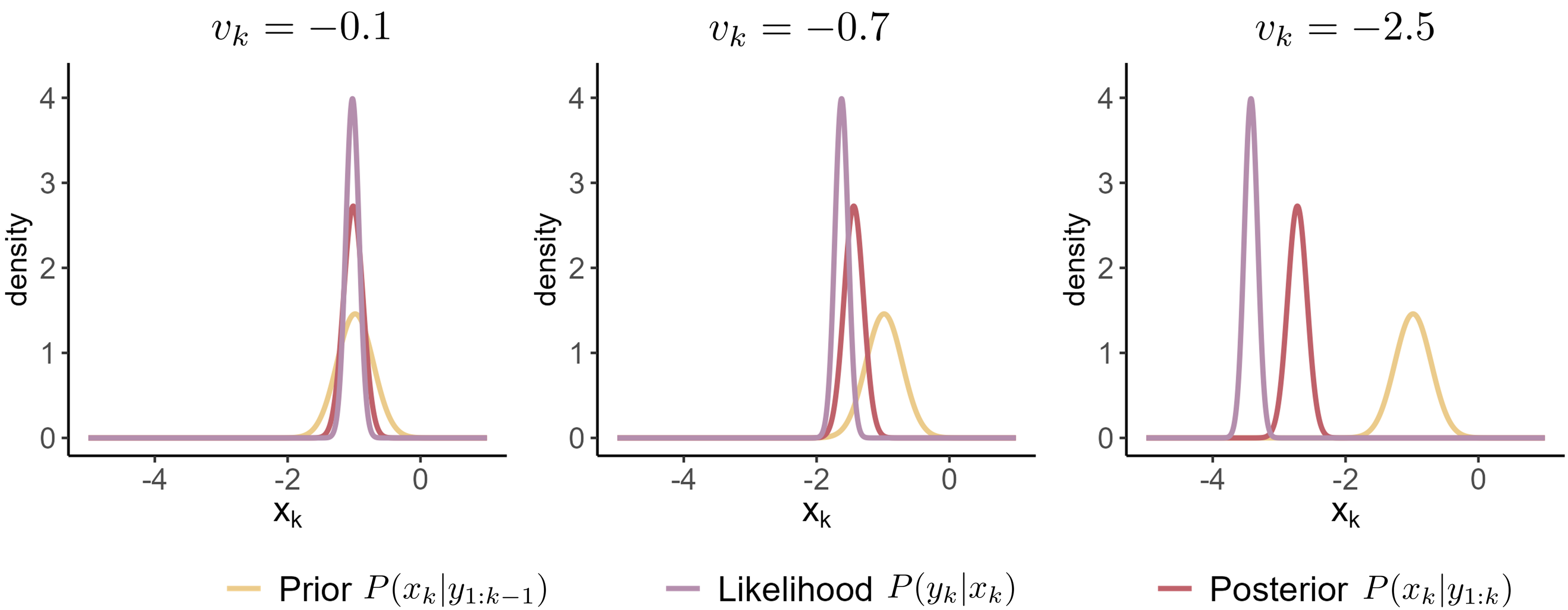}} \\
    \subfloat[Proposed AL-based filter]{\includegraphics[width=1\linewidth]{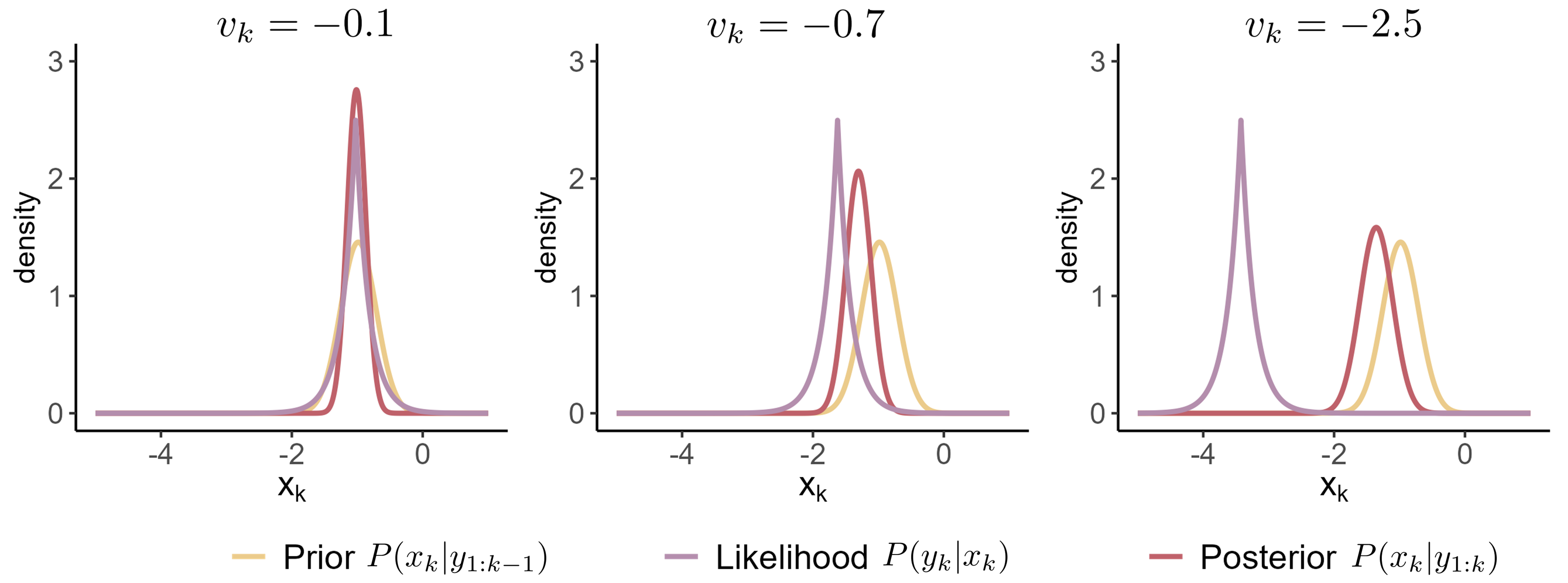}}
    \caption{Sensitivity to noise $v_k$ for different filters.}
    \label{fig: illustration of robustness-three bells}
\end{figure}

\subsubsection{VB inference} Although the mean-field approximation is used, our proposed AL-based filter exhibits similar robustness properties. Fig. \ref{fig: illustration of robustness-three bells} compares the Kalman filter and the AL-based filter under different noise levels $v_k$. The subfigures show the prior, likelihood, and posterior distributions for different values of $v_k$. When $\left|v_{k}\right|$ is small, both the Kalman filter and our AL-based filter produce comparable results, assigning similar weights to the prior and likelihood. However, when $\left|v_{k}\right|$ is large, the Kalman filter posterior becomes strongly biased by the likelihood, leading to poor performance in the presence of outliers. In contrast, our AL-based filter remains robust, effectively mitigating the influence of outliers and maintaining reliable state estimation.

\section{Parameter Estimation\label{sec:Parameter Estimation}}
In the previous sections, we focused on solving the filtering and smoothing problem for our proposed state-space model \ref{eq: state space problem}, assuming that the model parameters $\boldsymbol{\theta}$ were known. In this section, we address the case when these parameters are unknown and discuss how to estimate them. We begin with a high-level overview of the algorithmic structure and methods employed, specifically highlighting the efficiency of the single-loop algorithm. Subsequently, we delve into the mathematical details. For simplicity, throughout this section, $\mathbf{Y}$, $\mathbf{X}$, and $\boldsymbol{\lambda}$ will denote $y_{1:T}$, $\mathbf{x}_{1:T}$, and $\lambda_{1:T}$, respectively.

\subsection{Algorithmic Structure}
The EM algorithm is a powerful method for parameter estimation in models with latent variables. In the context of state-space models \cite{Huang2016identification, Neri2021}, EM typically employs a double-loop structure, where each outer loop contains an E-step and an M-step. However, we demonstrate that an efficient single-loop algorithm can be developed within the variational framework, potentially offering computational advantages while maintaining estimation accuracy.

\subsubsection{Conventional double-loop framework}
In robust state-space modeling, the E-step calculates the expected log-likelihood of latent variables given observed data and current parameters. Our variational approach extends this by estimating expectations for both $\mathbf{X}$ and $\boldsymbol{\lambda}$, denoted as $\mathbb{E}_{\mathbf{X},\boldsymbol{\lambda}\vert \mathbf{Y}, \boldsymbol{\theta}^i}[P(\mathbf{X},\boldsymbol{\lambda},\mathbf{Y})]$, computed via the smoothing algorithm (Algorithm \ref{al:smoother}). The M-step then maximizes this expected log-likelihood with respect to $\boldsymbol{\theta}$.

\subsubsection{Variational perspective on the double-loop framework}
We now present an equivalent perspective that reframes the EM algorithm in the variational context using the evidence lower bound (ELBO). This viewpoint is mathematically equivalent to the conventional framework but offers insights into the optimization process. The log-likelihood is defined as:
\begin{equation}
\log P\left(\mathbf{Y}|\boldsymbol{\theta}\right)=\mathcal{L}\left(q_x,q_{\lambda}, \boldsymbol{\theta}\right)+\text{KL}\left(q_x q_{\lambda}\parallel P\right),
\end{equation}
where $\text{KL}(q_x q_{\lambda}\parallel P) \geq 0$ represents the KL divergence between the variational approximations and the true posterior. The ELBO $\mathcal{L}(q_x,q_{\lambda}, \boldsymbol{\theta})$ provides a lower bound on $\log P(\mathbf{Y}|\boldsymbol{\theta})$, expressed as:
\begin{equation}
\begin{aligned}\mathcal{L}\left(q_{x},q_{\lambda},\boldsymbol{\theta}\right)= & \mathbb{E}_{q_{x},q_{\lambda}}\left[\log P\left(\mathbf{Y},\mathbf{X},\boldsymbol{\lambda}|\boldsymbol{\theta}\right)\right]\\
 & -\mathbb{E}_{q_{x}}\left[\log q_{x}\left(\mathbf{X}\right)\right]-\mathbb{E}_{q_{\lambda}}\left[\log q_{\lambda}\left(\boldsymbol{\lambda}\right)\right].
\label{eq:lower bound}
\end{aligned}
\end{equation}

In this formulation, the E-step is cast as an optimization problem of the ELBO over the variational distributions:
\begin{equation}
\text{E-step:}\quad\left(q_{x}^{i+1},q_{\lambda}^{i+1}\right)=\underset{q_{x},\:q_{\lambda}}{\mathsf{argmax}}\:\mathcal{L}\left(q_{x},q_{\lambda},\boldsymbol{\theta}^{i}\right).
\label{eq: general E-step}
\end{equation}
This employs alternating optimization, updating $q_x$ and $q_\lambda$ sequentially until convergence, aligning with Algorithm \ref{al:smoother}. Importantly, this optimization is equivalent to computing the expectations in the conventional E-step, as both approaches aim to find the best approximation of the posterior distribution given the current parameters.

The M-step maximizes the ELBO with respect to $\boldsymbol{\theta}$, using the updated distributions from the E-step:
\begin{equation}
\text{M-step:}\quad\boldsymbol{\theta}^{i+1}=\underset{\boldsymbol{\theta}}{\mathsf{argmax}}\:\mathcal{L}\left(q_{x}^{i+1},q_{\lambda}^{i+1},\boldsymbol{\theta}\right),
\label{eq: general M-step}
\end{equation}
Here, alternating optimization is also employed, where each inner iteration involves sequentially updating individual model parameters. This updating sequence is completed once for all parameters and repeated until convergence is achieved. This approach is equivalent to maximizing the expected log-likelihood in the conventional framework \cite{bishop2006pattern}.

\begin{figure}
    \centering
    \includegraphics[width=0.9\linewidth]{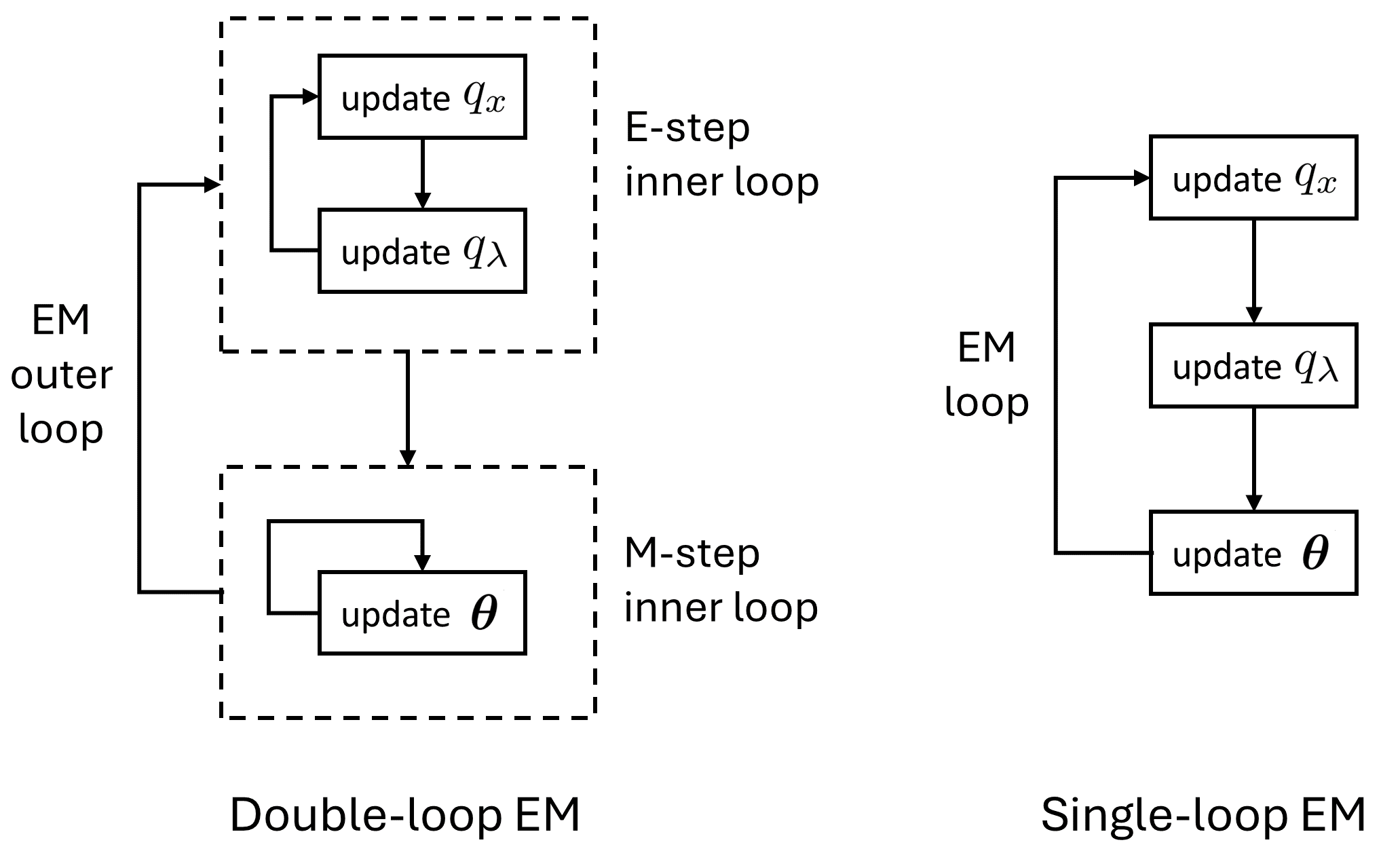}
    \caption{Workflow of the double-loop EM and the proposed single-loop EM.}
    \label{fig: workflow algorithm}
\end{figure}

The workflow of this double-loop framework is illustrated in Fig. \ref{fig: workflow algorithm}. Despite typically achieving convergence within a few outer iterations, the primary computational burden arises from the numerous, computationally intensive forward-backward iterations required within each E-step, which significantly impact the overall computational load.

\subsubsection{Efficient single-loop framework}
To address the computational inefficiencies of the conventional double-loop framework, we introduce a novel single-loop algorithm as illustrated in Fig. \ref{fig: workflow algorithm}. This streamlined approach simplifies the computational process by performing a single update of $q_x$ and $q_\lambda$ during the E-step, and a single update of $\boldsymbol{\theta}$ during the M-step. Each outer iteration now efficiently progresses with just one forward-backward pass, significantly reducing the total computational effort needed to achieve convergence.

The effectiveness of the single-loop framework stems from the fact that a single update in each of the E-step and M-step secures an increase in $\mathcal{L}$. Therefore, alternating these updates ensures a consistent and monotonic increase in $\mathcal{L}$. While our approach shares principles with the Generalized EM method used in traditional non-variational EM \cite{neal1998view}, it distinctively simplifies both the E-step and M-step, significantly reducing computational workload in variational contexts.

\begin{figure}
    \centering
    \includegraphics[width=0.45\linewidth]{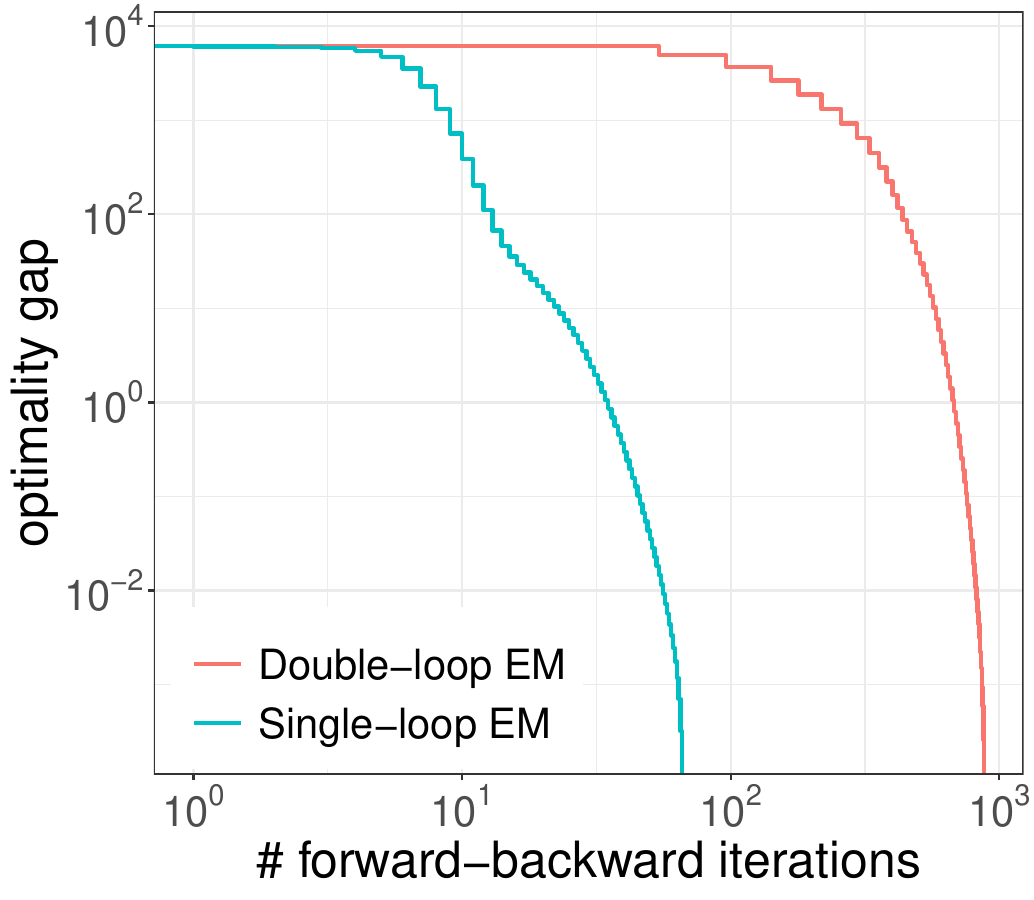}
    \includegraphics[width=0.45\linewidth]{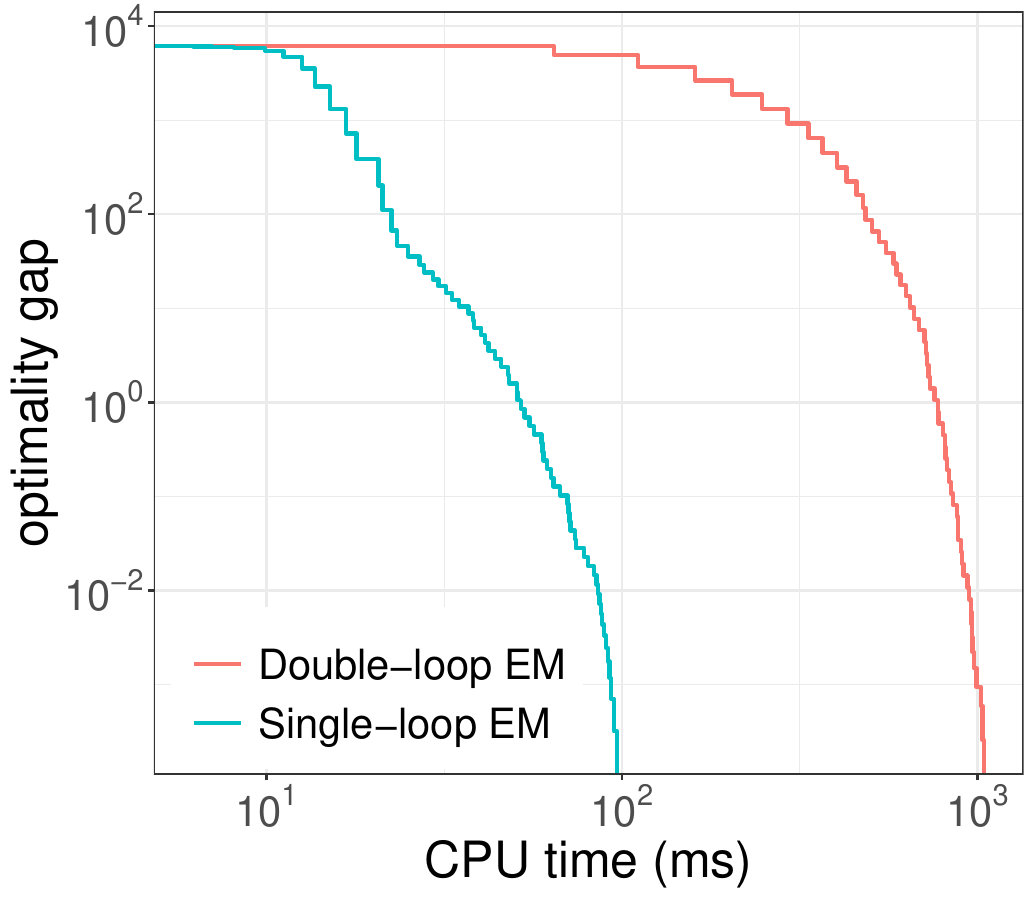}
    \caption{Convergence of the double-loop EM and the proposed single-loop EM.}
    \label{fig: convergence}
\end{figure}
Fig. \ref{fig: convergence} provides a comparison of the optimality gap between the double-loop algorithm and the single-loop algorithm in a sample experiment. Remarkably, the single-loop design demonstrates a significant reduction in the total number of forward-backward iterations, resulting in approximately 10 times faster convergence speed.

\begin{remark}
The single-loop framework is unsuitable for other approximate inference techniques like Expectation Propagation (EP) \cite{Neri2021}, which lacks the mechanism to guarantee such consistent increases in the variational lower bound.
\end{remark}

\subsection{Mathematical Details}
This subsection provides a detailed explanation of the single-loop parameter estimation algorithm.

\subsubsection{E-step}
The E-step of the single-loop algorithm updates $q_x$ and $q_\lambda$ as follows:
\begin{equation}
\begin{aligned}q_{x}^{i+1}= & \underset{q_{x}}{\mathsf{argmax}}\:\mathcal{L}\left(q_{x},q_{\lambda}^{i},\boldsymbol{\theta}^{i}\right),\\
q_{\lambda}^{i+1}= & \underset{q_{\lambda}}{\mathsf{argmax}}\:\mathcal{L}\left(q_{x}^{i+1},q_{\lambda},\boldsymbol{\theta}^{i}\right).
\end{aligned}
\end{equation}
These updates are achieved through a single forward-backward pass, specifically one iteration of the smoother as outlined in Algorithm \ref{al:smoother}. The aim of the E-step is to refine estimates of the latent variables using the current parameters.

\subsubsection{M-step}
The M-step focuses on updating the parameters, distinguishing between the concave and non-concave aspects of the objective function. For parameters excluding $p$, which affect $\mathcal{L}$ in a concave manner, we update each parameter $\theta$ sequentially by finding the value $\theta = \theta^{i+1}$ that satisfies the first-order optimality condition with closed-form solutions:
\begin{equation}
\partial \mathcal{L} / \partial \theta \big|_{\theta = \theta^{i+1}} = 0.
\label{eq:update theta/p}
\end{equation}

The parameter $p$, however, presents a unique challenge due to its contribution to the non-concavity of the simplified objective function $h(p)$ expressed as:
\begin{equation}
\begin{aligned}h\left(p\right)= & \frac{N}{2}\log p(1-p)-\alpha_1\frac{(\frac{1}{2}-p)^{2}}{2p(1-p)}\\
 & -\alpha_2p-\alpha_3p(1-p).
\end{aligned}
\label{eq:p EM}
\end{equation}
Here, $\alpha_1, \alpha_3> 0$ and $\alpha_2 \in \mathbb{R}$ are constants. Directly maximizing $h(p)$ within the constraint $p \in (0,1)$ is challenging due to the non-concave term $-\alpha_3p(1-p)$. To address this, we employ the Majorization-Minimization (MM) technique\cite{sun2016majorization, scutari2018parallel}, constructing a concave surrogate function $\tilde{h}$ using its first-order Taylor approximation:
\begin{equation}
\begin{aligned}\tilde{h}\left(p;p^{i}\right)= & \frac{N}{2}\log p(1-p)-\alpha_1\frac{(\frac{1}{2}-p)^{2}}{2p(1-p)}\\
 & -\alpha_2p-\alpha_3p(1-2p^{i}).
\end{aligned}
\label{eq:surrogate for p}
\end{equation}
We then solve the following optimization problem to update $p$:
\begin{equation}
p^{i+1}=\underset{p\in(0,1)}{\mathsf{argmax}}\:\tilde{h}\left(p;p^{i}\right).
\label{eq: surrogate prob for p}
\end{equation}
The analytical solution involves finding the roots of $d \tilde{h}(p; p^{i})/dp = 0$, and it can be proven that at least one root exists within the interval $(0,1)$. 
The MM approach not only guarantees that $\mathcal{L}(p^{i+1}) \geq \mathcal{L}(p^i)$, ensuring a non-decreasing sequence in $\mathcal{L}$, but also facilitates an efficient computation by providing a closed-form update for the parameter $p$. Detailed derivations of the M-step are provided in the supplementary material.

\subsubsection{Proposed algorithm}
Then, the complete single-loop algorithm is provided in Algorithm \ref{al:learn}.
\begin{algorithm}[t]
    \caption{Single-loop EM algorithm for parameter estimation in the state-space model \ref{eq: state space problem}\protect\label{al:learn}} 
    \begin{flushleft} 
    \textbf{Input:} Initial guess of parameter $\boldsymbol{\theta}^0$
    \end{flushleft}
    \begin{algorithmic}[1]
    \State Let $i = 0$
    \Repeat
        \baselineskip=14pt  
        \State Update $q_{x}^{i+1}$ according to Equation \eqref{eq: smoother update of q_x}
        \State Update $q_{\lambda}^{i+1}$ according to Equation \eqref{eq: smoother update of q_lambda}
        \State Update $\boldsymbol{\theta}^{i+1}$ according to Equations (61)-(74) in the supplementary material.
        \State Let $i = i+1$
    \Until{convergence}
    \end{algorithmic}
    \begin{flushleft}
    \textbf{Output:} Estimated parameters $\boldsymbol{\theta}^{\star}=\boldsymbol{\theta}^{i}$.
    \end{flushleft}
\end{algorithm}

\section{Experiments}\label{sec: experiments}

This study evaluates the performance of our proposed filter through three experiments. Firstly, we show that AL distribution is a good alternative to different well-known distribution with skewness or heavy tails, which makes it a generally applicable noise assumption in the state-space settings. Secondly, we show that the proposed AL-based filter outperforms the existing robust filters. Thirdly, we show the practical usage of the proposed model  in financial modeling.

\subsection{Synthetic Experiment I: Applicability of AL Distribution}
In this experiment, we assess the applicability of the proposed methods by applying it to synthetic data generated under different noise assumptions. Our goal is to demonstrate that the AL distribution is a suitable and versatile noise assumption in state-space settings. 

\subsubsection{Experimental settings}
We adopt a setup similar to \cite{Neri2021}, considering a two-state system with the following state transition matrix:
\begin{equation}
\mathbf{A}=\left[\begin{array}{cc}
\cos\left(0.2\pi\right) & \sin\left(0.2\pi\right)\\
-\sin\left(0.2\pi\right) & \cos\left(0.2\pi\right)
\end{array}\right].
\end{equation}
The observation matrix $\mathbf{C}\in \mathcal{R}^{N \times 2}$ is generated by sampling from a zero-mean isotropic Gaussian distribution, and its rows are then normalized such that $|C_{i,1} + C_{i,2}|=1$, for all $i=1,2,\dots,N$. The state noise covariance matrix $\mathbf{Q}$ is set to be $\mathbf{I}\sigma^2$, where $\sigma^2=0.05$. The constants terms $\mathbf{b}$ is set to $\mathbf{0}$.

We categorize the ground truth noise distributions into four groups:
\begin{itemize}
    \item No heavy tail or skewness: Gaussian distribution.    
    \item Skewed but not heavy-tailed: Skew-normal distribution \cite{azzalini2013skew}.
    \item Heavy-tailed but not skewed: Laplace and Student's t distributions \cite{Etemad2018locationT}.
    \item Heavy-tailed and skewed: AL and generalized hyperbolic skew-t distributions \cite{aas2006generalized}.
\end{itemize}
For each experiment, synthetic data with a length of $T=3000$ will be generated using the described linear state-space dynamics. The ground truth measurement noise $\mathbf{v}_t$ will be selected from the above distributions. We estimate the model parameters of different robust methods using the first $T\in[1,1500]$ data points and evaluate their smoothing performance on the remaining $T\in[1500,3000]$ data points.

\subsubsection{Benchmark methods}
Our proposed AL-Smoother is compared with the following benchmark methods, which utilize different noise assumptions and algorithms, for comparison.
\begin{itemize}
    \item True particle smoother (True-PS) in the Python package $\mathsf{particles}$ \cite{chopin2020particles}: Particle Filter assuming the true noise distributions, considered as the gold standard.
    \item Student's t particle smoother (T-PS) in the MATLAB function $\mathsf{bssm}$ \cite{matlab2022econometrics}: Particle Smoother assuming Student's t noise.
    \item Skew-normal particle smoother (SN-PS) in the MATLAB function $\mathsf{bssm}$ \cite{matlab2022econometrics}: Particle Smoother assuming skew-normal noise.
    \item Gaussian Kalman smoother (Gaussian-KS) in the R package $\mathsf{MARSS}$ \cite{holmes2012marss}: Kalman smoother assuming Gaussian noise.
\end{itemize}
Unlike the other methods, the True-PS operates with known parameters and does not require parameter estimation.

\subsubsection{Univariate results}

\begin{figure}
    \centering
    \includegraphics[width=1\linewidth]{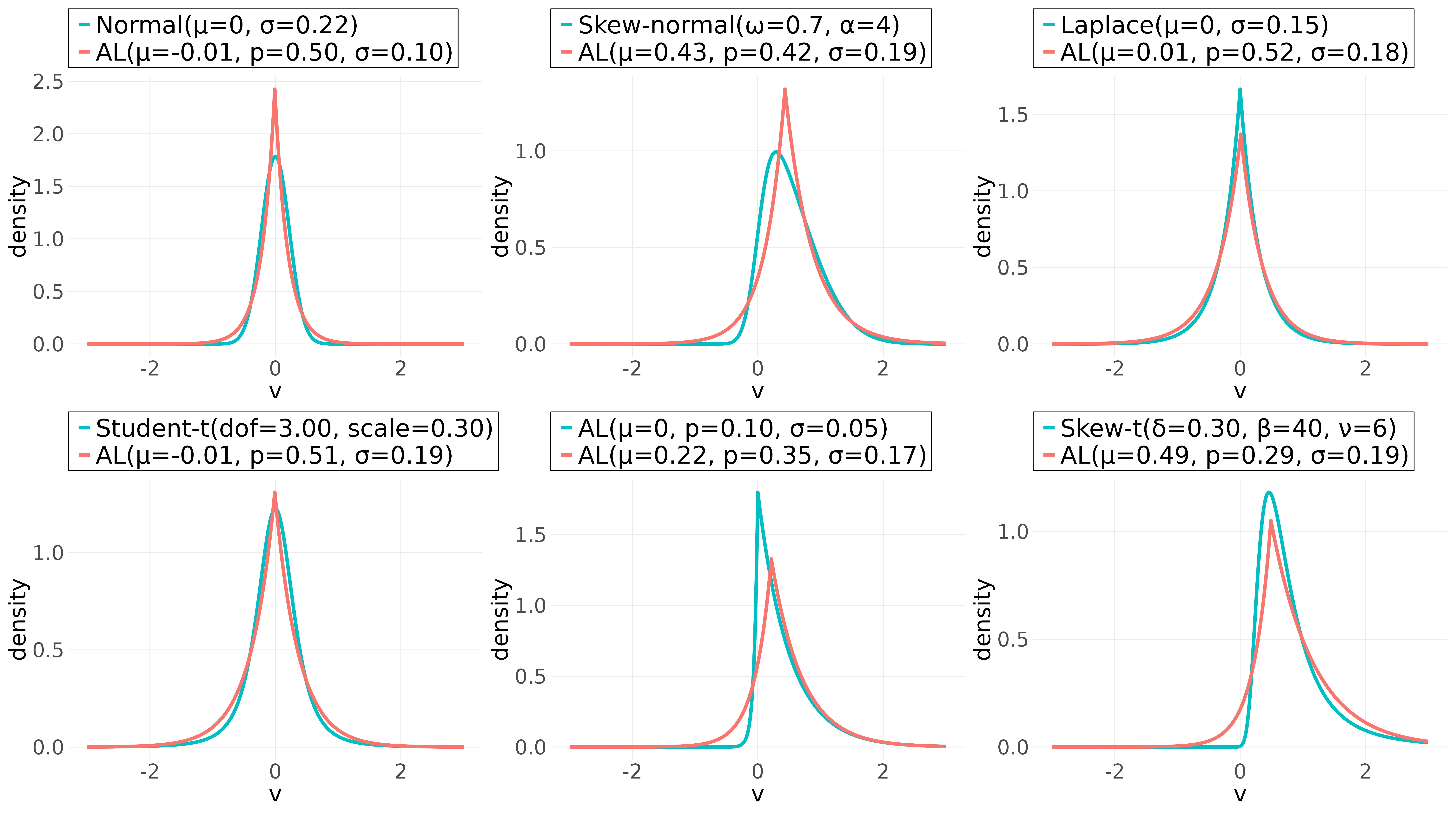}
    \caption{PDF of different ground truth noise distributions (blue) and the learned AL-distributed noise from AL-Smoother (red).}
    \label{fig:exp1_distribution}
\end{figure}

To start, we conduct a univariate experiment with $N = 1$. Fig. \ref{fig:exp1_distribution} displays subplots showing the PDF of the ground truth measurement noise alongside the fitted AL distributed noise obtained through our proposed EM algorithm. These plots demonstrate that the fitted AL distribution consistently captures the skewness and heavy-tailed nature of various noise assumptions.

\begin{figure}
    \centering
    \includegraphics[width=1\linewidth]{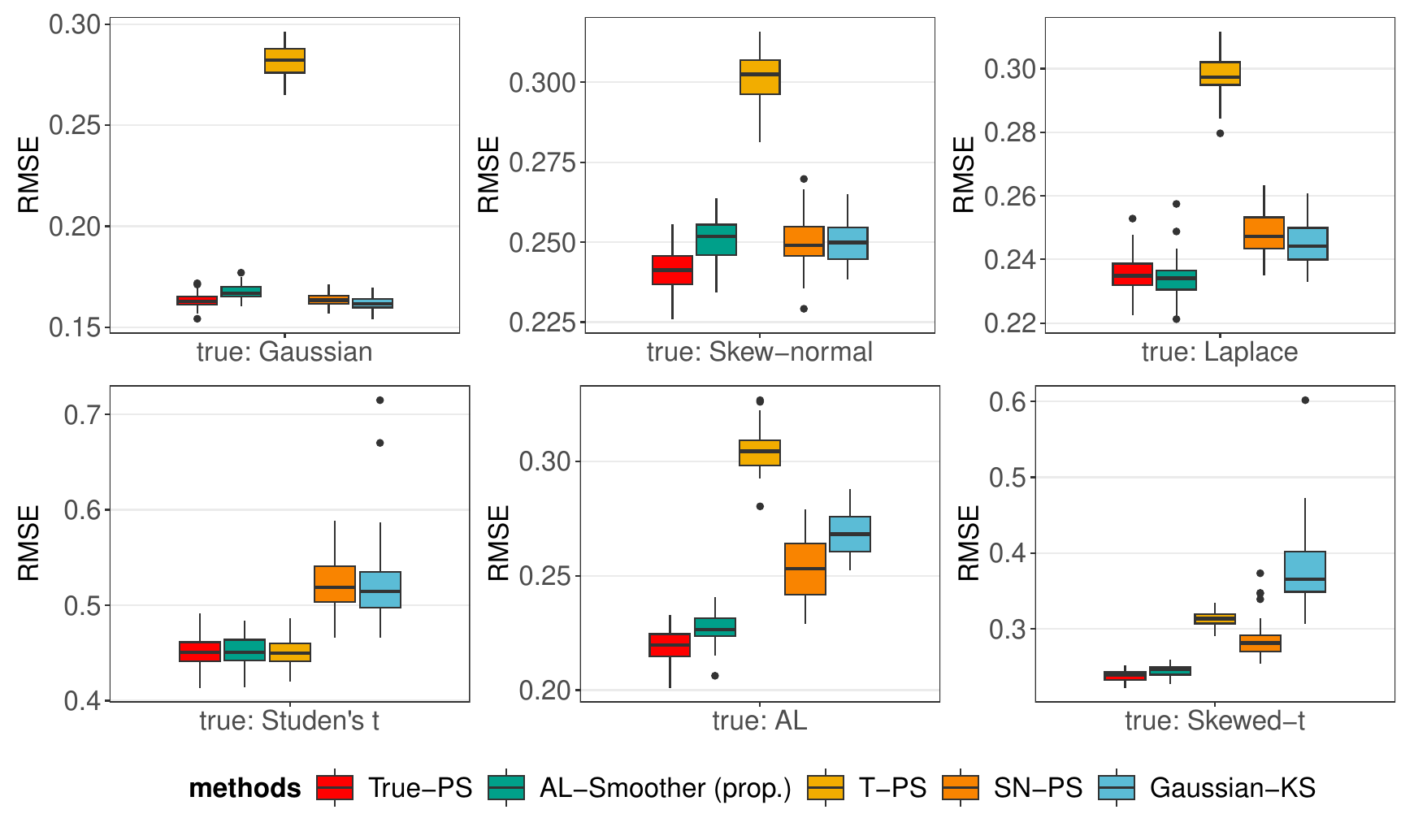}
    \caption{RMSE of the smoothed state estimate using various methods under different ground truth noise distributions.}
    \label{fig:exp1_rmse_independent}
\end{figure}

We compare the proposed AL-Smoother with the benchmark methods in estimating the smoothed hidden state under different assumptions of $\mathbf{v}_t$. The performance, evaluated using the root mean square error (RMSE), is shown in Fig. \ref{fig:exp1_rmse_independent}. The proposed method consistently exhibits comparable performance to the gold standard, True-PS. In contrast, the other benchmark methods (T-PS, SN-PS, and Gaussian-KS), which rely on specific noise assumptions, generally perform well only when the assumption matches the true distribution. Notably, the widely used T-PS encounters significant performance degradation when the ground truth noise assumption deviates from its own distribution, primarily due to the inaccuracy in estimating the degree of freedom of the noise.

\subsubsection{Multivariate results}

\begin{figure}
    \centering
    \includegraphics[width=0.8\linewidth]{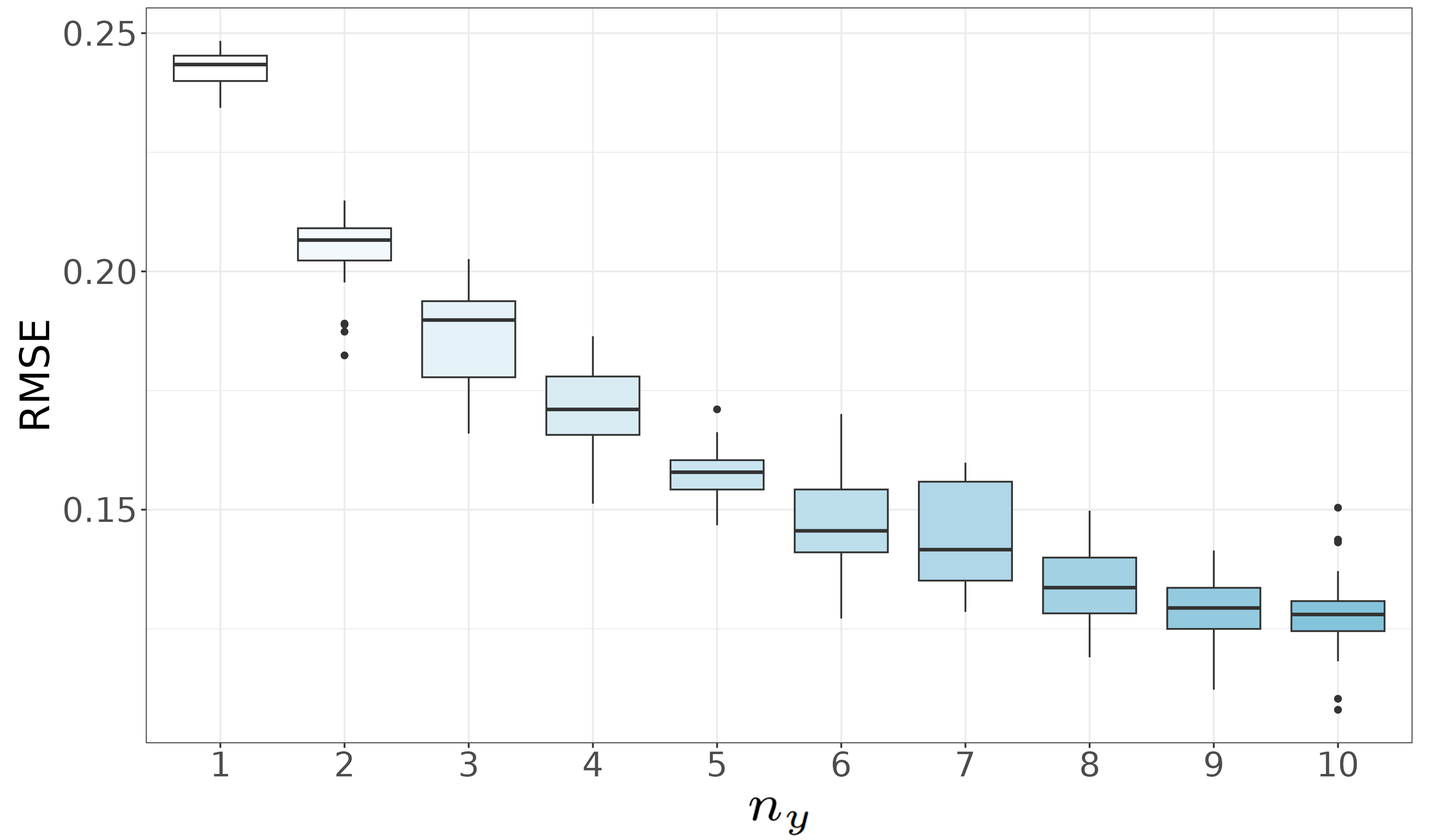}
    \caption{RMSE of the smoothed state estimates using AL-Smoother against the observation dimension $n_y$ under skew-t noise assumption.}
    \label{fig:exp1_multi_rmse}
\end{figure}

In addition to supporting univariate analysis, the proposed method also accommodates multiple observation sequences. We consider the case when $n_y$ sensors have independent skew-t noise with $\delta=0.3$, $\beta = 40$ and $\nu = 6$ \cite{aas2006generalized}. Fig. \ref{fig:exp1_multi_rmse} shows the RMSE of smoothed state estimate from our proposed method as $n_y$ increases from 1 to 10. The results reveal that the RMSE error monotonically decreases with increasing $n_y$, highlighting the efficiency of the proposed method in multivariate parameter learning and smoothing.

\subsection{Synthetic Experiment II: Robustness to Outliers}
We evaluate the robustness of our proposed filtering method to outliers by comparing it against two groups of approaches: existing robust filters with diverse structures and variants of MFVB methods on the same state-space model \ref{eq: state space problem}. Our objective is to demonstrate the superior performance of our method in handling outliers compared to these alternatives.

\subsubsection{Experimental Settings}
We utilize a simplified random-walk model with $n_x = n_y = 1$, and coefficients $A = C = 1$ and $b = 0$. The state noise covariance matrix $Q$ is set to 0.05. To simulate a typical sensor noise scenario, we model the measurement noise as a contaminated Gaussian noise pattern, similar to the setting in \cite{huang2019novel}, which consists of two components:
\begin{equation}
v_{t}\sim\begin{cases}
\mathcal{N}(0,r) & \text{with probability}\ 0.8;\\
\mathcal{N}(2,50r) & \text{with probability}\ 0.2,
\end{cases}
\end{equation}
where $r = 0.01$. The primary component, accounting for 80\% of the data, models the standard output from a well-functioning sensor with zero mean and low variance. The secondary component, representing 20\%, introduces a bias with a mean of 2 and a variance significantly higher, simulating sensor miscalibration or environmental interference.

We generate 20 training sets and 100 testing sets, each containing 1000 data points. We assess the accuracy of the filtered state values $\hat{x}_{k\vert k}$ against the true state values $x_k$, using RMSE and Maximum Error (EMax). EMax is specifically defined as:
\begin{equation}
    \text{EMax}(\hat{x}_{k\vert k}) = \underset{1\leq k \leq T}{\mathsf{max}}\: \vert\hat{x}_{k\vert k} -x_k\vert.
\end{equation}

\subsubsection{Comparison I: Existing robust filters}
We compare both filters we proposed, which both utilize the noise distribution specified by $\mathcal{AL}(0,p,\sigma)$, where the mean $\mu$ is set to zero.
They are compared to several existing robust filters:
\begin{itemize}
    \item Laplace filter (LF), using a percentile parameter of $p=0.5$ based on our proposed Fast AL-Filter, with noise distribution $\mathcal{AL}(0,0.5,\sigma_l)$.
    \item Skew-t filter (STF), using a skew-t noise model $\mathcal{ST}(0,\sigma_{st}^2,\delta_{st},\nu_{st})$, as proposed in \cite{Nurminen2015}.
    \item Student's t filter (TF), assuming Student's t noise $\mathcal{T}(0,\sigma_t^2, \nu_t)$, as proposed in \cite{agamennoni2011outlier}.
    \item Huberisation based Kalman filter (HKF), using Gaussian noise $\mathcal{N}(0,\sigma_h^2)$ and applying the Huber loss function with threshold $\xi$ to the post-fit residual, as proposed in \cite{ruckdeschel2014robust}.
    \item Adaptive filter (ADF), using Gaussian noise $\mathcal{N}(0,\sigma_a^2)$ and adjusting noise covariance based on a look-back window $N_{\text{win}}$ in Equation \eqref{eq: adaptive R in mohamed}, as proposed in \cite{mohamed1999adaptive}.
    \item Particle filter with model selection (MSPF) \cite{martino2017cooperative}, where three noise models are used, consisting of the true noise distribution, $\mathcal{AL}(0, 0.3, 0.2)$ and $\mathcal{AL}(0, 0.1, 0.2)$.
\end{itemize}
The implementations of all methods are carried out in R and executed on 2.10GHz Intel Core i7-12700 machines. 

For Exact AL-Filter, Fast AL-Filter and LF, parameters $\sigma_l = 0.289$, $\sigma = 0.162$, and $p = 0.22$ are learned using our proposed Algorithm \ref{al:learn}. For STF and TF, the degrees of freedom are set to $\nu_t = \nu_{st} = 4$, a common choice for robust filters \cite{Nurminen2018,Nurminen2015,lange1989robust}. Lacking specific
learning algorithms for other parameters in STF and TF, we adapt a variational EM algorithm, achieving $\sigma_t^2 = 0.155$, $\sigma_{st}^2 = 0.240$, and $\delta_{st} = 0.129$. The threshold $\xi$ for HKF is determined using the method described in \cite{ruckdeschel2014robust}. All other parameters for HKF and ADF are optimized through an exhaustive grid search, resulting in $\sigma_h^2 = 0.15$, $\sigma_a^2 = 0.5$, and $N_{\text{win}} = 30$. Note that the MSPF method is provided with the true noise distribution, serving as the golden standard.

\begin{figure}[]
\begin{centering}
\subfloat[Metric: RMSE]{\includegraphics[width=0.9\linewidth]{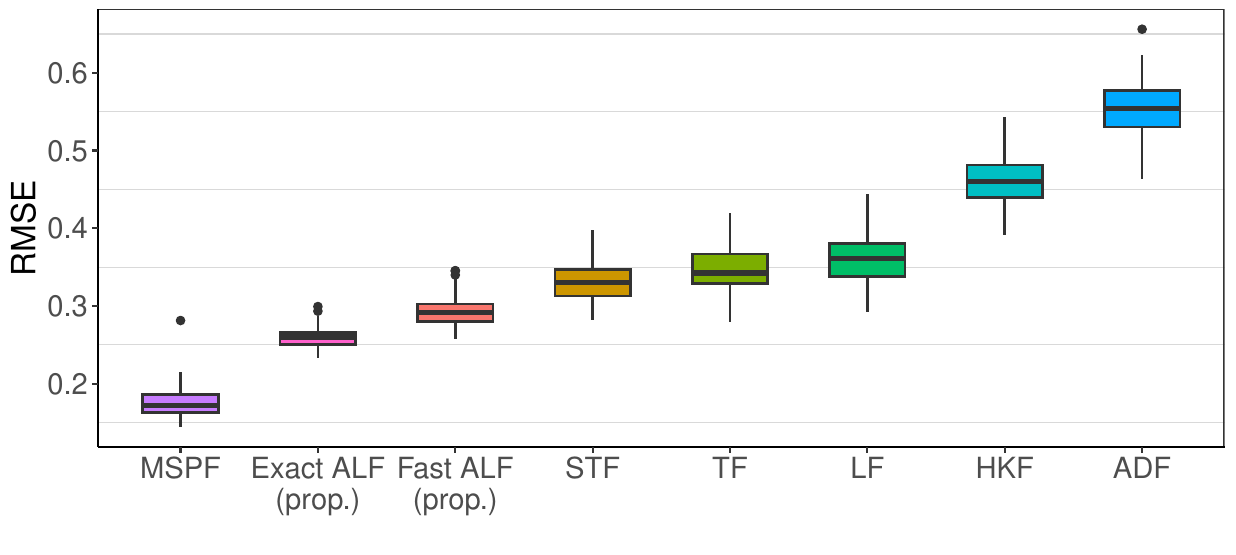}}  \hfill
\subfloat[Metric: EMax]{\includegraphics[width=0.9\linewidth]{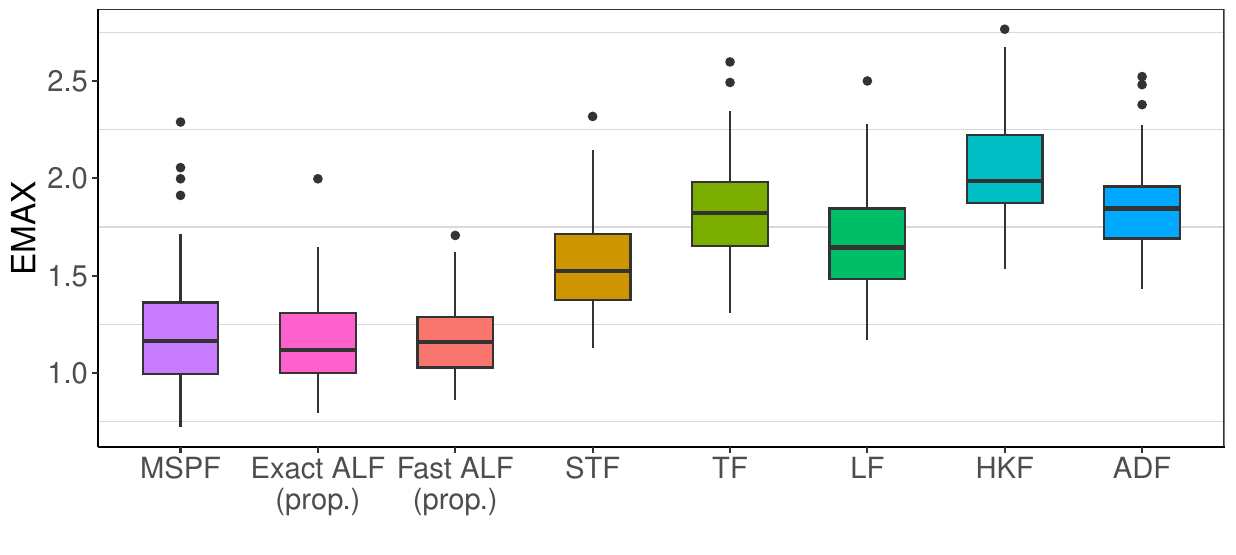}}
\par\end{centering}
\caption{Comparison with existing robust filters.}
\label{fig:exp2-rmseemax-filter}
\end{figure}
\begin{table}
\centering
\begin{tabular}{|c|c|c|c|}
\hline 
\textbf{MSPF} & \textbf{Exact ALF (prop.) } & \textbf{Fast ALF (prop.)} & \textbf{STF}\tabularnewline
\hline 
47.61 & 234.36 & 0.78 & 2.28\tabularnewline
\hline 
\textbf{TF} & \textbf{LF} & \textbf{HKF} & \textbf{ADF}\tabularnewline
\hline 
0.70 & 0.72 & 0.07 & 0.05\tabularnewline
\hline 
\end{tabular}
\caption{Average CPU time of proposed methods and benchmarks (unit: second)\label{table: exp2}}
\end{table}

Fig. \ref{fig:exp2-rmseemax-filter} presents the results for the different methods. Table \ref{table: exp2} shows the running time of the methods. The MSPF achieves excellent RMSE performance as the true noise model is included. However, this approach requires pre-specification of all possible noise models, and its performance is influenced by the appropriateness of the candidate model collection.  In contrast, our Fast AL-Filter learns parameters adaptively from data while maintaining comparable
worst-case performance (EMax) and achieving substantially better computational efficiency. 

For the Exact AL-Filter, while it maintains theoretical rigor, it requires much longer running time compared to the Fast AL-Filter. Importantly, both methods achieve comparable robustness performance in terms of RMSE and EMax metrics, validating that the Fast AL-Filter's approximations do not significantly compromise filtering accuracy in practice. This makes the proposed Fast AL-Filter particularly valuable in scenarios where computational resources are constrained and robustness to outliers is needed.

While both the AL and skew-t distributions capture heavy tails and skewness, the Fast AL-Filter demonstrates more consistent and solid performance than STF, TF, and LF, which do not account for skewness, perform relatively worse. Non-probabilistic methods, i.e., HKF and ADF, generally show weaker performance compared to probabilistic methods, likely due to their sensitivity to hyperparameter choices.

\subsubsection{Comparison II: Variants of MFVB methods on the state-space model \ref{eq: state space problem}}
Our proposed filters effectively addresses \ref{eq: state space problem} by employing a fixed parameter approach while performing variational inference on $\mathbf{X}$ and $\boldsymbol{\lambda}$.
Another set of MFVB methods applicable to \ref{eq: state space problem} treats $\boldsymbol{\theta}$ as a random variable, extending MFVB across all variables ($\mathbf{X}$, $\boldsymbol{\lambda}$, and $\boldsymbol{\theta}$). A key concern is that performance is highly sensitive to the prior chosen for $\boldsymbol{\theta}$, which is typically non-informative. To evaluate potential improvements, we introduce an informed prior based on our Algorithm \ref{al:learn}. Furthermore, while only approximate MFVB is typically discussed \cite{huang2017novel}, we have developed an exact MFVB for a more comprehensive comparison. The variants under comparison include:
\begin{itemize}
    \item Informed-prior exact MFVB (IPE)
    \item Informed-prior approximate MFVB (IPA)
    \item Non-informed-prior exact MFVB (NIPE)
\end{itemize}
\begin{figure}[]
\begin{centering}
\subfloat[Metric: RMSE]{\includegraphics[width=0.48\linewidth]{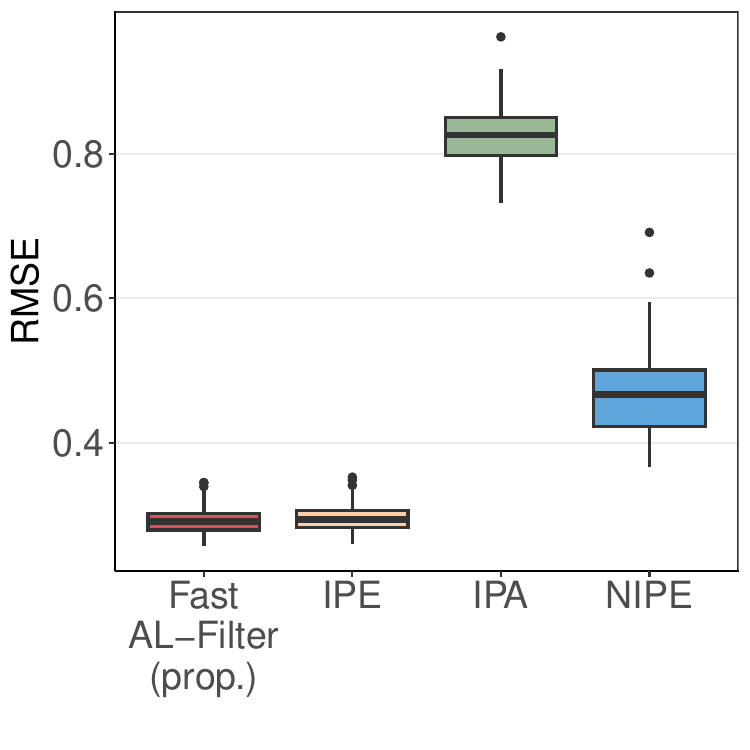}} \hfill
\subfloat[Metric: EMax]{\includegraphics[width=0.48\linewidth]{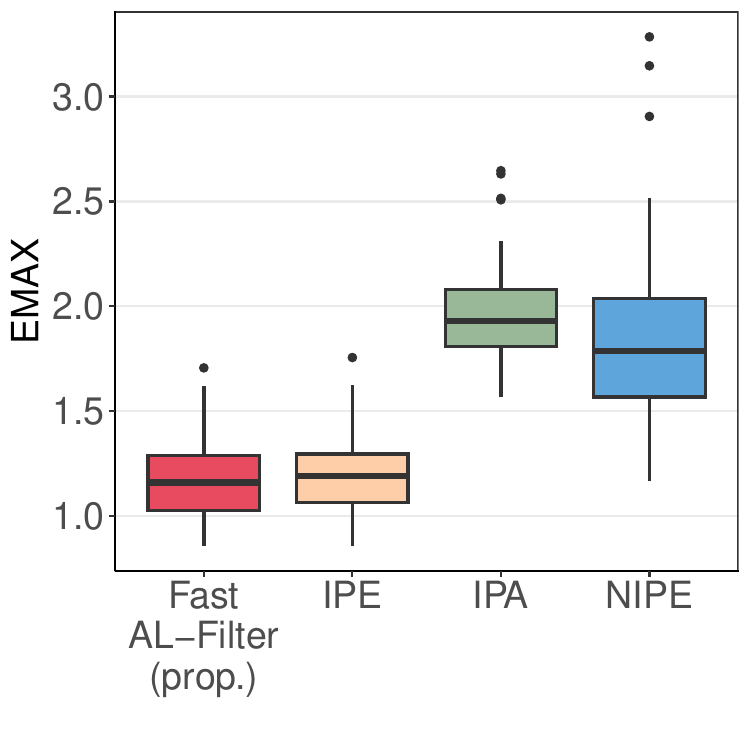}}
\par\end{centering}
\caption{Comparison with variants of MFVB methods on the state-space model \ref{eq: state space problem}.}
\label{fig:exp2-rmseemax-Huang}
\end{figure}
Fig. \ref{fig:exp2-rmseemax-Huang} shows the performance differences among the variants of MFVB methods on \ref{eq: state space problem}. The proposed Fast AL-Filter, which fixes $\boldsymbol{\theta}$, demonstrates superior performance. 
When $\boldsymbol{\theta}$ is treated as a random variable with an informed prior, IPE yields performance closely matching our Fast AL-Filter. However, significant performance degradation occurs with both IPA and NIPE. These findings underscore that fixing $\boldsymbol{\theta}$ after learning provides more robust outcomes compared to methods that treat $\boldsymbol{\theta}$ as a random variable, where the choice of prior and the availability of closed-form updates critically influence performance.

\subsection{Empirical Experiment: Stochastic Volatility Model}
The third experiment showcases a practical application of our proposed method in the stochastic volatility (SV) model.

\subsubsection{State-space reformulation of SV}
A SV model describes the dynamics of asset prices by introducing the concept of volatility as an unobserved random variable $\sigma_k$ that evolves over time \cite{taylor1994modeling}. A simple stationary SV model is given by
\begin{equation}
\begin{aligned}h_{k}= & \phi h_{k-1}+\gamma+\eta_{k}\\
y_{k}= & \exp\left(h_{k}/2\right)z_{k}.
\end{aligned}
\label{eq: exp3-sv-origin}
\end{equation}
Here, $h_k=\ln{(\sigma_k^2)}$, $y_k$ represents the demeaned return of an asset, and $\eta_k\sim \mathcal{N}(0, \sigma_\eta)$ and $z_{k}\sim\mathcal{N}(0, 1)$ are independent white noise processes. However, the SV model (\ref{eq: exp3-sv-origin}) is a nonlinear state-space model with an intractable likelihood, making the estimation of parameters $\boldsymbol{\theta} = \left\{ \phi,\gamma, \sigma_\eta \right\}$ and volatility series challenging. One common solution is the computationally intensive MCMC approach (e.g., R package $\mathsf{stochvol}$ \cite{hosszejni2019stochvol}).

Alternatively, to facilitate parameter estimation, the model (\ref{eq: exp3-sv-origin}) can be reformulated into a linear state-space model:
\begin{equation}
\begin{aligned}h_{k}= & \phi h_{k-1}+\gamma+\eta_{k}\\
\ln\left(y_{k}^{2}\right)= & h_{k}+\ln\left(z_{k}^{2}\right).
\end{aligned}
\label{eq: exp3-sv-reform}
\end{equation}
Since the observation noise $\ln(z_{k}^{2})$ follows a log chi-square distribution, an approximation is required. One approach, suggested by \cite{ruiz1994quasi} and \cite{harvey1994multivariate}, is to approximate the observation noise as Gaussian with known mean and variance, via Maximum Likelihood Estimation (MLE), namely ${v}_{k}\sim \mathcal{N}(-1.27, \pi^2 /2)$. This leads to the following reformulation:
\begin{equation}
\begin{aligned}h_{k}= & \phi h_{k-1}+\gamma+\eta_{k}\\
\ln\left(y_{k}^{2}\right)= & h_{k}+{v}_{k}.
\end{aligned}
\label{eq: exp3-sv-gaussian}
\end{equation}
The parameters $\boldsymbol{\theta}$ can be estimated using the EM algorithm for linear Gaussian state-space models. The volatility series is then recovered from the smoothed state estimates as $\hat{\sigma}_k = [\exp(\hat{h}_{k|T})]^{1/2}, k=1,\dots,T$. However, since $\ln(z_{k}^{2})$ is skewed and heavy-tailed, the Gaussian approximation is often insufficient.

\begin{figure}
    \centering
    \includegraphics[width=0.8\linewidth]{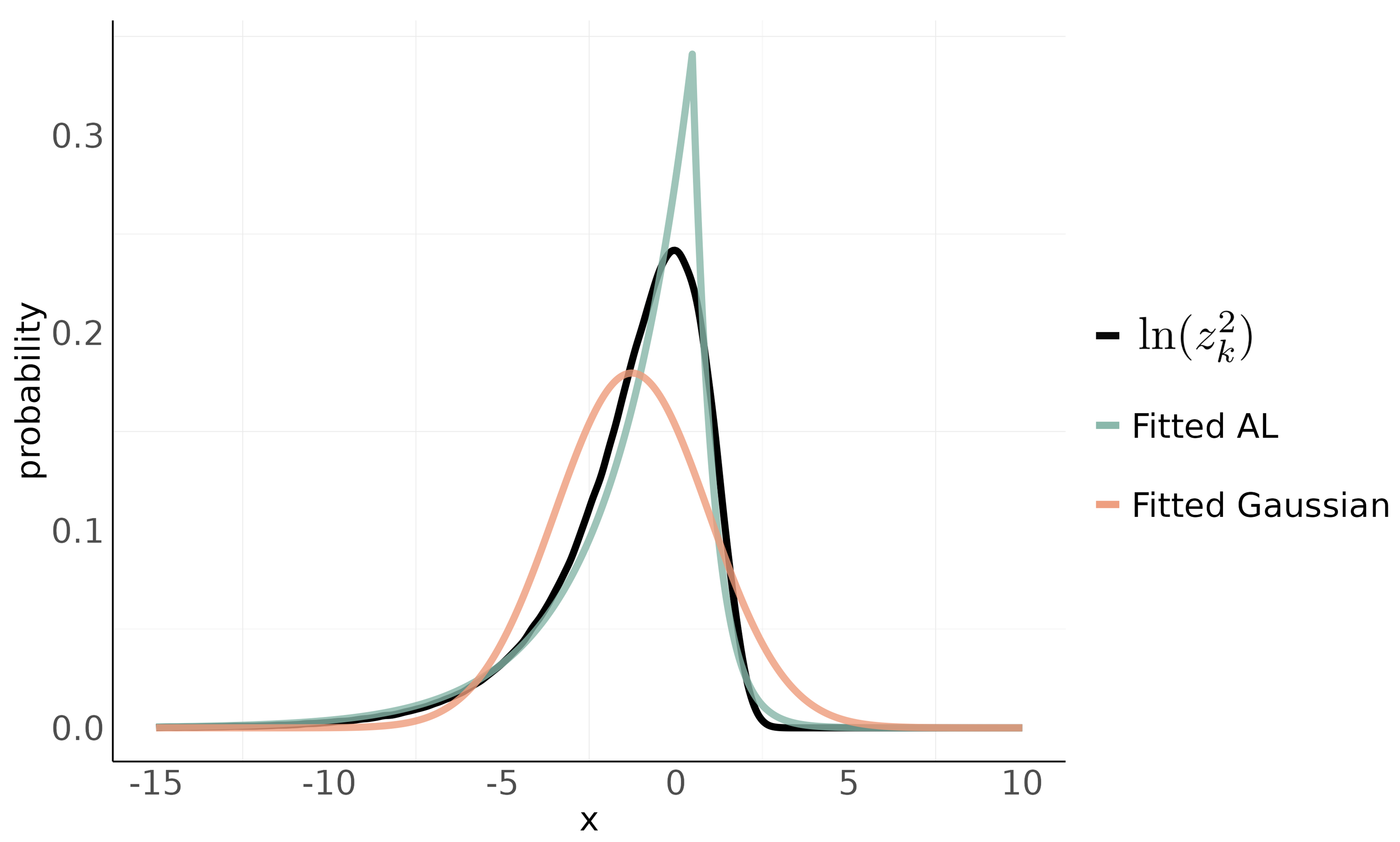}
    \caption{MLE fits of market noise $\ln(z_{k}^{2})$ using AL and Gaussian distributions.}
    \label{fig: exp3_fit_noise}
\end{figure}

\begin{table}
    \centering
    \begin{tabular}{|c|c|c|c|c|}
    \hline 
     & Mean & Variance & Skewness & Kurtosis\tabularnewline
    \hline 
    $\ln(z_{k}^{2})$ & -1.27 & 4.93 & -16.81 & 170.47\tabularnewline
    \hline 
    $\ensuremath{\mathcal{AL}(0.48,\ 0.8,\ 0.47)}$ & -1.28 & 5.84 & -25.33 & 283.68\tabularnewline
    \hline 
    $\ensuremath{\mathcal{N}(-1.27, \ \pi^{2}/2)}$ & -1.27 & 4.93 & 0 & 73.05\tabularnewline
    \hline 
    \end{tabular}
    \caption{Empirical moments of $\ln(z_{k}^{2})$ and MLE fitted AL and Gaussian distributions.}
    \label{tab: exp3_moments}
\end{table}

AL distribution offers an alternative approximation of $\ln(z_{k}^{2})$. One approximation is ${v}_{k}\sim \mathcal{AL}(\mu=0.48,p=0.8,\sigma=0.47)$ according to MLE \cite{yu2005three}. This approximation provides a more accurate representation, matching the higher moments of $\ln(z_{k}^{2})$, as indicted in Fig. \ref{fig: exp3_fit_noise} and Table \ref{tab: exp3_moments}. After modeling $v_k$ as an AL-distributed variable, the reformulated state-space model fits into \ref{eq: state space problem} and can be solved with our proposed methods.

\subsubsection{Experimental Settings} To evaluate the performance of different SV models,
we perform experiments on historical daily price time series data of S\&P 500 stocks from January 2010 to December 2023. It includes first estimating unknown parameters $\boldsymbol{\theta}$ and then estimating the volatility series. Our proposed method, denoted as Fast AL-Filter, is compared with the following methods:
\begin{itemize}
    \item MCMC-based SV model in R package $\mathsf{stochvol}$ \cite{hosszejni2019stochvol} with default 10,000 draws, considered as the gold standard.
    \item Kalman filter approach by \cite{ruiz1994quasi} and \cite{harvey1994multivariate}.
\end{itemize}

\subsubsection{Results} 

\begin{figure}
    \centering
    \includegraphics[height=3.2cm]{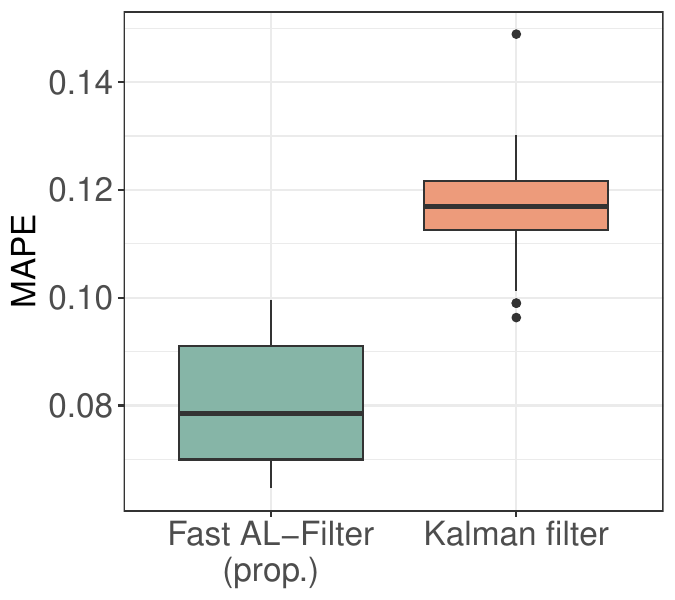}
    \caption{MAPE comparison of the Fast AL-Filter and Kalman filter, using results from $\mathsf{stochvol}$ as a benchmark.}
    \label{fig: exp3_MAPE}
\end{figure}
We assess the accuracy of estimating the stochastic volatility series $\sigma_k$ using both the Kalman filter and the proposed Fast AL-Filter. The accuracy is quantified through the mean absolute percentage error (MAPE), taking the estimates from the $\mathsf{stochvol}$ package as the benchmark. As depicted in Fig. \ref{fig: exp3_MAPE}, Fast AL-Filter demonstrates a reduction in error of approximately 30\% compared to the Kalman filter, a result of its refined reformulation of the SV model.
\begin{figure}
    \begin{centering}
    \captionsetup[subfloat]{justification=centering} 
    \subfloat[CPU time\label{fig: exp3_time}]{\includegraphics[height=3.2cm]{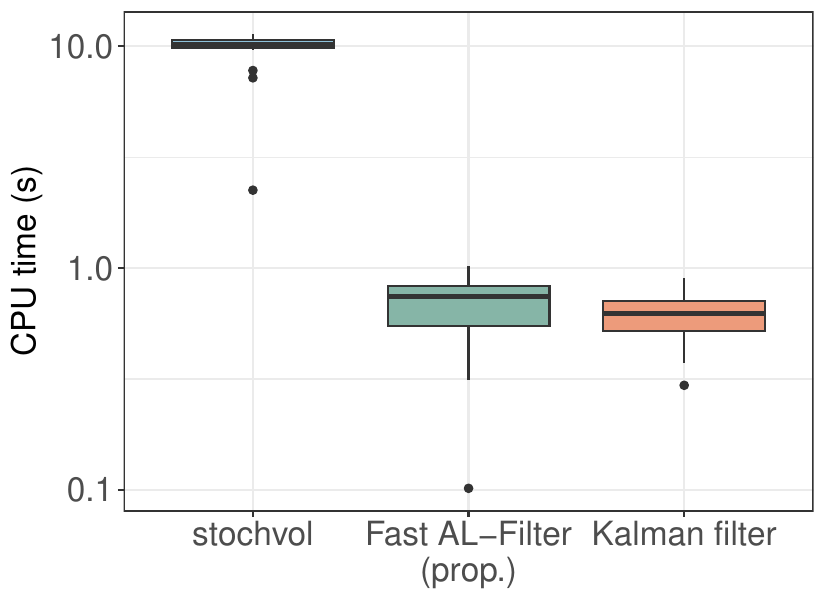}}
    \hspace{1.5em}%
    \subfloat[forward-backward iterations\label{fig: exp3_iteration}]{\includegraphics[height=3.2cm]{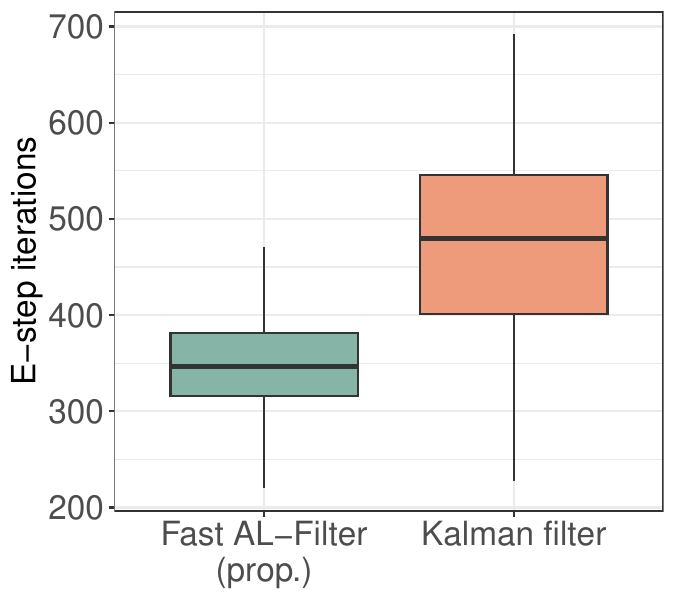}}
    \par\end{centering}
    \caption{Computational cost of different methods.}
\end{figure}

Fig. \ref{fig: exp3_time} displays the CPU time for each method, encompassing both parameter estimation and smoothing processes. The proposed Fast AL-Filter requires fewer forward-backward iterations during parameter estimation than the Kalman filter, as detailed in Fig. \ref{fig: exp3_iteration}, contributing to its comparable computational efficiency. In contrast, the $\mathsf{stochvol}$ package, utilizing an MCMC approach, requires an order of magnitude more time due to the extensive number of draws necessary to achieve reliable results.

In conclusion, our analysis demonstrates that the proposed Fast AL-Filter not only aligns closely with the MCMC-based approach in terms of estimation accuracy but also significantly reduces computational time, making its efficiency comparable to that of the traditional Kalman filter.

\section{Conclusion}\label{sec: conclusion}
This study presents a robust extension of traditional state-space models utilizing the AL distribution, meticulously crafted to adeptly manage outliers characterized by skewness and heavy tails. By integrating an efficient VB inference algorithm with a novel single-loop parameter estimation strategy, we have significantly enhanced the computational efficiency of the filtering, smoothing, and parameter estimation processes. The robustness of our approach is thoroughly elucidated through a detailed interpretation of adaptive filtering techniques and Bayesian robustness, emphasizing how our model evolves uniquely compared to other methods. Our experiments demonstrate the model's superior performance in both synthetic and empirical settings, showcasing its adaptability to diverse noise environments that typically challenge existing methods, which often depend on specific noise characteristics or require extensive manual tuning. These findings establish the proposed AL-based model and algorithms as valuable tools for applications burdened by non-Gaussian noise and data anomalies, confirming their practical relevance and effectiveness.

\appendices

\section{Derivation of the smoother update}\label{appendix:update_x_and_lambda}
When updating $\mathbf{x}_{1:T}$ as per Equation \eqref{eq: smoother update of q_x}, we focus solely on terms relevant to $\mathbf{x}_{1:T}$:
\begin{equation}
\begin{aligned}
\label{eq: update smoother}
        & \log q_x^{j+1}(\mathbf{x}_{1:T}) \\=& \log P(\mathbf{x}_1) + \sum_{k = 2}^T \log P(\mathbf{x}_{k}\vert \mathbf{x}_{k-1})  \\
         & +\sum_{k = 1}^T \mathbb{E}_{q_{\lambda}^{j}}\left[\log P(y_k\vert \mathbf{x}_k,\lambda_k)\right] + \mathbb{E}_{q_{\lambda}^{j}}\left[\log P(\lambda_k)\right]\\
         = & -\frac{1}{2}\sum_{k=2}^{N}\left(\mathbf{x}_{k}-\mathbf{A}\mathbf{x}_{k-1}\right)^{\intercal}\mathbf{Q}^{-1}\left(\mathbf{x}_{k}-\mathbf{A}\mathbf{x}_{k-1}\right)\\
 & -\frac{1}{2}\sum_{k=1}^{N}\left(y_{k}-\mathbf{C}\mathbf{x}_{k}-\frac{\left(\frac{1}{2}-p\right)\sigma}{\mathbb{E}_{q_{\lambda}^{j}}[\lambda_{k}]p\left(1-p\right)}\right)^{2}\frac{\mathbb{E}_{q_{\lambda}^{j}}[\lambda_{k}]p\left(1-p\right)}{\sigma^{2}}\\
 & -\frac{1}{2}\left(\mathbf{x}_{1}-\boldsymbol{\pi}_{1}\right)^{\intercal}\boldsymbol{\Sigma}_{1}^{-1}\left(\mathbf{x}_{1}-\boldsymbol{\pi}_{1}\right)+\text{const}
\end{aligned}
\end{equation}
When updating $\lambda_{1:T}$ as per Equation \eqref{eq: smoother update of q_lambda}, we focus solely on terms relevant to $\lambda_{1:T}$:
\begin{equation}
\begin{aligned}\log q^{j+1}_{\lambda}\left(\lambda_{k}\right)= & \mathbb{E}_{q_x}\left[\log P\left(\mathbf{x}_{1:T},\lambda_{1:T},y_{1:T}\right)\right]+\text{const}\\
= & \sum_{k = 1}^T-\frac{\lambda_{k}}{2\sigma^{2}}\mathbb{E}_{q_x^{j+1}}\left[\left(y_{k}-\mathbf{C}\mathbf{x}_{k}-\mu\right)^{2}\right]p(1-p)\\
 & -\frac{1}{\lambda_{k}8p(1-p)}-\frac{3}{2}\log\lambda_{k}+\text{const}.
\end{aligned}
\end{equation}
This derivation confirms that $\lambda_k$ follows an inverse Gaussian distribution.

\section{Derivation of the filter update}\label{appendix:filter update_x_and_lambda}
Given the conjugate Gaussian prior, the variational posterior $q^{j+1}_{x}(\mathbf{x}_{k})$ takes a Gaussian form:
\begin{equation}
\label{eq:q_x closed-form}
q^{j+1}_{x}\left(\mathbf{x}_{k}\right)=\mathcal{N}\left(\mathbf{x}_{k}\vert\hat{\mathbf{x}}_{k\vert k},\Sigma_{k\vert k}\right),
\end{equation}
where 
\begin{equation}
\label{eq:filter-K}
    \mathbf{K}^{j+1}_k = \boldsymbol{\Sigma}_{k|k-1} \mathbf{C}\mystrut^\intercal\left(\mathbf{C}\boldsymbol{\Sigma}_{k|k-1}\mathbf{C}\mystrut^\intercal + r_k^{j}\right)^{-1}
\end{equation}
\begin{equation}
\label{eq:filter-x}
    \hat{\mathbf{x}}^{j+1}_{k|k} = \hat{\mathbf{x}}_{k|k-1} + \mathbf{K}^{j+1}_k\left(y_{k}- \mathbf{C}\hat{\mathbf{x}}_{k|k-1} - m_k^{j}\right)
\end{equation}
\begin{equation}
\label{eq:filter-Sigma}
       \boldsymbol{\Sigma}^{j+1}_{k|k} = \boldsymbol{\Sigma}_{k|k-1} - \mathbf{K}^{j+1}_k\mathbf{C}\boldsymbol{\Sigma}_{k|k-1}
\end{equation}
with $r_k^{j} = \displaystyle{\frac{\sigma^2}{\mathbb{E}_{q_\lambda^{j}}[\lambda_{k}] p(1-p)}}$, $m_k^{j} = \mu + \displaystyle{\frac{(\frac{1}{2}-p)\sigma}{\mathbb{E}_{q_\lambda^{j}}[\lambda_{k}] p(1-p)}}$.

The derivation of $q_\lambda(\lambda_k)$ is similar to the Appendix \ref{appendix:update_x_and_lambda}, and thus is omitted here. The required mean value 
\begin{equation}
\label{eq:lambda mean}
    \mathbb{E}_{q_\lambda^{j+1}}[\lambda_k] = \frac{\sigma}{2p(1-p)\sqrt{\mathbb{E}_{q_x^{j+1}}[(y_k - \mathbf{Cx}_k- \mu)^2]}}.
\end{equation}
\section{Derivation of Equation \eqref{eq:bayesian robust}}\label{appendix:simple model}
The derivation is similar to Theorem 1 in \cite{pericchi1992exact} with minor modifications. Consider a prior $P(x_k\vert y_{1:k-1})$ and a likelihood $P(y_k \vert x_k)$. We begin with:
\begin{equation}
\begin{split}
     & \mathbb{E}\left[x_k\vert y_k\right] - \hat{x}_{k\vert k-1} \\&= \frac{\int (x_k - \hat{x}_{k \vert k-1})P(x_k\vert y_{1:k-1})P(y_k \vert x_k) d x_k}{P(y_k\vert y_{1:k-1})}.
\end{split}
\label{eq: appx B-first}
\end{equation}
Let us define:
\begin{equation}
    s(x_k;x) = \frac{1}{\sqrt{2\pi\Sigma_{k\vert k-1}}}\exp\left((x_k-x)^2\Sigma_{k\vert k-1}^{-1}\right),
\end{equation}
\begin{equation}
    g_k(x) = \int s(x_k;x)P(y_k\vert x_k) dx_k,
\end{equation}
where $s(x_k;\hat{x}_{k\vert k-1}) = P(x_k\vert y_{1:k-1})$ and $ g_k(\hat{x}_{k\vert k-1}) = P(y_k\vert y_{1:k-1})$. Using these definitions, we can rewrite Equation \eqref{eq: appx B-first} as:
\begin{equation}
\begin{split}
     &g_k(\hat{x}_{k\vert k-1})(\mathbb{E}\left[x_k\vert y_k\right] - \hat{x}_{k\vert k-1}) \\&
     = \int (x_k - \hat{x}_{k \vert k-1})P(x_k\vert y_{1:k-1})P(y_k \vert x_k) d x_k \\ &
     = \Sigma_{k\vert k-1}\int\left.\frac{\partial s(x_k;x)P(y_k \vert x_k)}{\partial x}\right\vert_{ x= \hat{x}_{k\vert k-1}}d x_k\\&
     = \Sigma_{k\vert k-1}\left.\frac{d g_k(x)}{dx}\right\vert_{ x= \hat{x}_{k\vert k-1}}.
\end{split}
\end{equation}
From this, we derive:
\begin{equation}
    \hat{x}_{k\vert k} = \mathbb{E}\left[x_k\vert y_k\right] = \hat{x}_{k\vert k-1} + \Sigma_{k\vert k-1}\left.\frac{d \log g_k(x)}{dx}\right\vert_{ x= \hat{x}_{k\vert k-1}}.
\end{equation}
The expression for $g_k(\cdot)$, derived through basic operations, is:
\begin{equation}
    \begin{split}
        g_k(x) = &u_k(x,p-1)\Psi\left(\frac{x-y_k}{\sqrt{\Sigma_{k\vert k-1} }}- \frac{(p-1)\sqrt{\Sigma_{k\vert k-1} }}{\sigma}\right) \\&
        + u_k(x,p)\Psi\left(\frac{y_k-x}{\sqrt{\Sigma_{k\vert k-1} }}- \frac{p\sqrt{\Sigma_{k\vert k-1} }}{\sigma}\right),
     \end{split}
    \end{equation}
where
    \begin{equation}
        u_k(x,z) = \frac{p(1-p)}{\sigma}\exp\left(\frac{z^2\Sigma_{k\vert k-1}}{2\sigma^2} + \frac{z(x-y_k)}{\sigma}\right),
        \end{equation}
and $\Psi(\cdot)$ denotes the cumulative distribution function of the standard Gaussian distribution.

\bibliographystyle{IEEEtran}

\bibliography{paper}
\clearpage
\onecolumn
\section*{\Large Supplementary Material}
In this supplementary material, we give detailed derivation of M-step.

Please note that the M-step derivation provided below is conducted in a univariate context ($ n_y = 1 $) to enhance clarity. The objection function of $i$-th M-step can be written as
\begin{equation}
\begin{aligned}
&\mathcal{L}\left(q_{x}^{i+1},q_{\lambda}^{i+1},\boldsymbol{\theta}\right)\\= & \mathbb{E}\left[\log P\left( \mathbf{x}_{1:T}, \lambda_{1:T},y_{1:T} \right)| y_{1:T};\boldsymbol{\theta}^{i-1}\right]\\
= & -\mathbb{E}_{1}-\mathbb{E}_{2}-\mathbb{E}_{3}-\frac{1}{2}\log\left|\boldsymbol{\Sigma}_{1}\right|-\frac{T-1}{2}\log\left|\mathbf{Q}\right|\\
 & -\sum_{k=1}^{T}\mathbb{E}\left[\frac{3}{2}\log \lambda_{k}+\frac{1}{2\lambda_{k}}\right]-T\log \sigma+\frac{T}{2}\log p\left(1-p\right)+\text{const},
\end{aligned}
\end{equation}
where the terms $\mathbb{E}_1$, $\mathbb{E}_2$, and $\mathbb{E}_3$ are defined as follows:
\begin{equation}
\begin{aligned}\mathbb{E}_{1}= & \mathbb{E}\left[\frac{1}{2}\sum_{k=2}^{T}\left(\mathbf{x}_{k}-\mathbf{A}\mathbf{x}_{k-1}-\mathbf{b}\right)^{\intercal}\mathbf{Q}^{-1}\left(\mathbf{x}_{k}-\mathbf{A}\mathbf{x}_{k-1}-\mathbf{b}\right)\right]\\
= & \frac{1}{2}\sum_{k=2}^{T}\left[\text{Tr}\left(\mathbf{Q}^{-1}\mathbf{P}_{k}\right)-\text{Tr}\left(\mathbf{Q}^{-1}\mathbf{A}\mathbf{P}_{k,k-1}^\intercal\right)\right.\\&-\left.\text{Tr}\left(\mathbf{Q}^{-1}\mathbf{P}_{k,k-1}\mathbf{A}^\intercal\right)
+\text{Tr}\left(\mathbf{A}^{\intercal}\mathbf{Q}^{-1}\mathbf{A}\mathbf{P}_{k-1}\right) \right.\\&+ \bigl.2\mathbf{b}^\intercal\mathbf{Q}^{-1}\mathbf{A}\hat{\mathbf{x}}_{k-1\vert T}  - 2\mathbf{b}^\intercal\mathbf{Q}^{-1}\hat{\mathbf{x}}_{k\vert T} \Bigl]+\frac{T-1}{2}\mathbf{b}^\intercal\mathbf{Q}^{-1}\mathbf{b},
\end{aligned}
\end{equation}

\begin{equation}
\begin{aligned}\mathbb{E}_{2}= & \mathbb{E}\left[\sum_{k=1}^{T}\left(y_{k}-\mathbf{C}\mathbf{x}_{k}-\mu-\frac{\left(\frac{1}{2}-p\right)\sigma}{\lambda_{k}p\left(1-p\right)}\right)^{2}\frac{\lambda_{k}p\left(1-p\right)}{2\sigma^{2}}\right]\\
= & \frac{1}{2}\sum_{k=1}^{T}\left[\frac{u_k}{r_k}+\left(\mathbf{C}\hat{\mathbf{x}}_{k\vert T}+\mu-y_{k}\right)\frac{1-2p}{\sigma}+\frac{(\frac{1}{2}-p)^2}{p(1-p)}\mathbb{E}\left[\frac{1}{\lambda_k}\right]\right],
\end{aligned}
\end{equation}

\begin{equation}
\begin{aligned}\mathbb{E}_{3}= & \mathbb{E}\left[\frac{1}{2}\left(\mathbf{x}_{1}-\boldsymbol{\pi}_{1}\right)^{\intercal}\boldsymbol{\Sigma}_{1}^{-1}\left(\mathbf{x}_{1}-\boldsymbol{\pi}_{1}\right)\right]\\
= & \frac{1}{2}\left[\text{Tr}\left(\boldsymbol{\Sigma}_{1}^{-1}\mathbf{P}_{1}\right)-2\boldsymbol{\pi}_{1}^{\intercal}\boldsymbol{\Sigma}_{1}^{-1}\hat{\mathbf{x}}_{1\vert T}+\boldsymbol{\pi}_{1}^{\intercal}\boldsymbol{\Sigma}_{1}^{-1}\boldsymbol{\pi}_{1}\right].
\end{aligned}
\end{equation}
Also, we require the following expectations and variables:
\begin{equation}
    \mathbf{P}_k = \mathbb{E}\left[\mathbf{x}_k\mathbf{x}_k^\intercal\vert y_{1:T},\boldsymbol{\theta}^{i-1} \right] = \boldsymbol{\Sigma}_{k\vert T} + \hat{\mathbf{x}}_{k\vert T}\hat{\mathbf{x}}_{k\vert T}^\intercal,
\end{equation}
\begin{equation}
       \mathbf{P}_{k,k-1} = \mathbb{E}\left[\mathbf{x}_k\mathbf{x}_{k-1}^\intercal\vert y_{1:T},\boldsymbol{\theta}^{i-1} \right] = \mathbf{L}_{k-1}\boldsymbol{\Sigma}_{k\vert T} + \hat{\mathbf{x}}_{k\vert T}\hat{\mathbf{x}}_{k-1\vert T}^\intercal,
\end{equation}
\begin{equation}
     \mathbb{E}\left[\lambda_k\right] = \frac{\sigma}{2p(1-p)\sqrt{u_k}},
\end{equation}
\begin{equation}
    \mathbb{E}\left[\frac{1}{\lambda_k}\right] = \frac{2p(1-p)\sqrt{u_k}}{\sigma} + 4p(1-p),
\end{equation}
and 
\begin{equation}
    u_k =\left(y_{k}-\mathbf{C}\hat{\mathbf{x}}_{k|T} -\mu\right)^2 + \mathbf{C}\boldsymbol{\Sigma}_{k\vert T}\mathbf{C}^\intercal,
\end{equation}
\begin{equation}
    r_k = \frac{\sigma^2}{p(1-p)\mathbb{E}\left[\lambda_k\right]}.
\end{equation}

Each parameter is updated sequentially using an alternating optimization approach, with each new value immediately applied in subsequent updates to enhance algorithm efficiency. For clarity, iteration superscripts are omitted in the following equations, assuming each parameter is at its latest update. Detailed equations are provided:
\begin{itemize}
    \item $\sigma$:
\begin{equation}
\label{eq:M-sigma}
    \sigma^{i+1}=\frac{1}{2}\left(\frac{\left(\frac{1}{2}-p\right)}{T}\sum_{k=1}^{T}\left(\mathbf{C}\hat{\mathbf{x}}_{k\vert T}+\mu-y_{k}\right)+\sqrt{\Delta}\right),
\end{equation}
where
\begin{equation}
    \Delta = \left(\frac{\left(\frac{1}{2}-p\right)}{T}\sum_{k=1}^{T}\left(\mathbf{C}\hat{\mathbf{x}}_{k\vert T}+\mu-y_{k}\right)\right)^2 +\frac{4}{T}\sum_{k=1}^{T}u_kp(1-p)\mathbb{E}\left[\lambda_k\right]
\end{equation}
\item $\mu$:

\begin{equation}
\label{eq:M-mu}
    \mu^{i+1} = -\left(\frac{T(1-2p)}{2\sigma}+\sum_{k =1}^T\frac{\mathbf{C}\hat{\mathbf{x}}_{k\vert T}-y_k}{r_k}\right)\left(\sum_{k =1}^N\frac{1}{r_k}\right)^{-1}
\end{equation}
\item $\boldsymbol{\pi}_1$
\begin{equation}
\label{eq:M-pi1}
    \boldsymbol{\pi}_1^{i + 1} = \hat{\mathbf{x}}_{1\vert T}
\end{equation}
\item $\boldsymbol{\Sigma}_1$:
\begin{equation}
\label{eq:M-Sigma1}
    \boldsymbol{\Sigma}_1^{i + 1} = \mathbf{P}_1 - \hat{\mathbf{x}}_{1\vert T}\hat{\mathbf{x}}_{1\vert T}^\intercal
\end{equation}
\item $\mathbf{A}$:
\begin{equation}
\mathbf{A}^{i+1}=\left(\sum_{k=2}^{T}\mathbf{P}_{k,k-1}-\mathbf{b}\hat{\mathbf{x}}_{k-1\vert T}^{\intercal}\right)\left(\sum_{k=2}^{T}\mathbf{P}_{k-1}\right)^{-1}
\end{equation}
\item $\mathbf{C}$:
\begin{equation}
\label{eq:M-C}
    \mathbf{C}^{i+1} = \left(\sum_{k =1}^T \frac{(y_k-\mu)\hat{\mathbf{x}}_{k\vert T}}{r_k} - \frac{1-2p}{2\sigma}\hat{\mathbf{x}}_{k\vert T}\right)\left(\sum_{k =1}^T \frac{\mathbf{P}_k}{r_k}\right)^{-1}
\end{equation}
\item $\mathbf{Q}$:

\begin{equation}
\label{eq:M-Q}
\begin{split}
     \mathbf{Q}^{i+1} =&\frac{1}{T-1} \sum_{k = 2}^T \Bigl(\mathbf{P}_k - \mathbf{P}_{k, k-1}\mathbf{A}^\intercal-\mathbf{A}\mathbf{P}_{k, k-1}^\intercal + \mathbf{A}\mathbf{P}_{k-1}\mathbf{A}^\intercal \Bigr.\\& -\mathbf{b}\hat{\mathbf{x}}_{k-1\vert T}^\intercal\mathbf{A}^\intercal   -\mathbf{A}\hat{\mathbf{x}}_{k-1\vert T}\mathbf{b}^\intercal \left.- \mathbf{b}\hat{\mathbf{x}}_{k-1\vert T}^\intercal-\hat{\mathbf{x}}_{k-1\vert T}\mathbf{b}^\intercal\right)
     + \mathbf{b}\mathbf{b}^\intercal
\end{split}
\end{equation}
\item $\mathbf{b}$:
\begin{equation}
\label{eq:M-b}
    \mathbf{b}^{i+1} =\frac{1}{T-1} \sum_{k = 2}^T\hat{\mathbf{x}}_{k\vert T} - \mathbf{A}\hat{\mathbf{x}}_{k-1\vert T}
\end{equation}
\item $p$: To avoid confusion, we denote $p^{i}$ as $p^{(i)}$ in this part.
We establish that
\begin{equation}
    \lim_{p \to 0^+}\nabla_p\tilde{h}(p; p^{(i)})\times\lim_{p \to 1^-}\nabla_p\tilde{h}(p; p^{(i)}) < 0,
\end{equation}
indicating opposite signs of the gradient at the endpoints of the interval $[0, 1]$. By the intermediate value theorem, there must exist at least one real root of the first derivative of $\tilde{h}(p; p^{(i)})$ expressed as
\begin{equation}
\tilde{h}\left(p;p^{(i)}\right)= \frac{N}{2}\log p(1-p)-\alpha_1\frac{(\frac{1}{2}-p)^{2}}{2p(1-p)}
 -\alpha_2p-\alpha_3p(1-2p^{(i)}),
\end{equation}
within the interval $[0, 1]$. Consequently, it is unnecessary to impose the constraint $0 < p < 1$ on the surrogate problem. The next iterate, $p^{(i+1)}$, is derived by solving the polynomial equation:
\begin{equation}
\label{eq:M-p}
    p^4 + \left(\frac{T}{\zeta_1} - 2\right) p^3 + \left(1-\frac{3T}{2\zeta_1}\right)p^2 +\frac{T + \zeta_2}{2\zeta_1}p -\frac{\zeta_2}{4\zeta_1}= 0,
\end{equation}
where 
\begin{equation}
    \zeta_1 = -\frac{1}{\sigma} \sum_{k = 1}^N(y_{k} - \mathbf{C}\hat{\mathbf{x}}_{k\vert T}-\mu)  - \frac{(1-2p^{(i)})}{2\sigma^2}\sum_{k = 1}^N\mathbb{E}[\lambda_{k}]u_k,
\end{equation}

\begin{equation}
\label{eq:end of Appendix M-step}
    \zeta_2 = - \frac{1}{2}\sum_{k = 1}^N\mathbb{E}\left[\frac{1}{\lambda_{k}}\right].
\end{equation}

\end{itemize}
\end{document}